\documentclass[10pt,journal,compsoc]{IEEEtran}
\usepackage{cite}
\usepackage{amsmath}
\usepackage{amsfonts}
\usepackage{color}
\usepackage{comment}
\usepackage{subfigure}
\usepackage{graphicx}
\usepackage{multirow}
\usepackage{algorithmic}
\usepackage{bm}
\usepackage{url}
\usepackage{booktabs}  
\usepackage{bbding}
\usepackage[linesnumbered, ruled,noend]{algorithm2e}
\newtheorem{Def}{Definition}

\begin{document}

\title{STMARL: A Spatio-Temporal Multi-Agent Reinforcement Learning Approach for Cooperative Traffic Light Control}

\author{ Yanan~Wang,
        Tong~Xu, Xin~Niu, Chang~Tan, \\Enhong Chen,~\IEEEmembership{Senior Member,~IEEE}
        and Hui Xiong,~\IEEEmembership{Fellow,~IEEE}
    
    \IEEEcompsocitemizethanks{
        \IEEEcompsocthanksitem Y.~Wang is with School of Computer Science, University
        of Science and Technology of China and 
Department of Management Information Systems, Eller College of Management, the University of Arizona.\protect\\
        E-mail: ynwwang@email.arizona.edu.
        \IEEEcompsocthanksitem T.~ Xu and E.~Chen are with Anhui Province Key Laboratory of Big
        Data Analysis and Application, School of Computer Science, University
        of Science and Technology of China, Hefei 230027, China.\protect\\
        E-mail: \{tongxu,chenehg\}@ustc.edu.cn.
        
        \IEEEcompsocthanksitem X.~Niu and C.~Tan are with iFLYTEK Research, IFLYTEK Co., Ltd, Hefei, Anhui 230088, China.
        \protect\\ Email: \{xinniu2,changtan2\}@iflytek.com.
        
        \IEEEcompsocthanksitem H. Xiong is with the Management Science and Information Systems Department, Rutgers Business School, Rutgers University, Newark, NJ 07102, USA.
        \protect\\ Email: hxiong@rutgers.edu.
        \IEEEcompsocthanksitem T.~ Xu and H. Xiong are corresponding authors.
    }
    
}

\markboth{}%
{Shell \MakeLowercase{\textit{et al.}}: Bare Demo of IEEEtran.cls for Computer Society Journals}

\IEEEtitleabstractindextext{%
\begin{abstract}
The development of intelligent traffic light control systems is essential for smart transportation management. While some efforts have been made to optimize the use of individual traffic lights in an isolated way, related studies have largely ignored the fact that the use of multi-intersection traffic lights is spatially influenced, as well as the temporal dependency of historical traffic status for current traffic light control. To that end, in this paper, we propose a novel Spatio-Temporal Multi-Agent Reinforcement Learning (STMARL) framework for effectively capturing the spatio-temporal dependency of multiple related traffic lights and control these traffic lights in a coordinating way. Specifically, we first construct the traffic light adjacency graph based on the spatial structure among traffic lights. Then, historical traffic records will be integrated with current traffic status via Recurrent Neural Network structure. Moreover, based on the temporally-dependent traffic information, we design a Graph Neural Network based model to represent relationships among multiple traffic lights, and the decision for each traffic light will be made in a distributed way by the deep Q-learning method.  Finally, the experimental results on both synthetic and real-world data have demonstrated the effectiveness of our STMARL framework, which also provides an insightful  understanding of the influence mechanism among multi-intersection traffic lights.
\end{abstract}

}

\maketitle

\IEEEpeerreviewmaketitle



\section{Introduction}
\IEEEPARstart{R}{ecent} years have witnessed \textcolor{black}{a sharp increase} in traffic congestion in most cities, which results in several negative effects like air pollution and economic losses. For instance, traffic congestion has caused the financial cost of $\$$305 billion in 2017 in the US, $\$$10 billion more than 2016~\cite{traficjamcost}. Along this line, optimal control of traffic lights has been widely used for reducing congestion in mobile environments\textcolor{black}{~\cite{xu2019exploring, sommer2010bidirectionally,wu2019block, cabannes2019regrets, li2020competitive, zhang2020semi}}. Traditionally, the control plan of traffic lights was pre-defined as fixed based on historical traffic data~\cite{miller1963settings, yin2016traffic}, or artificially regulated by officers based on current traffic status~\cite{cools2013self}. However, these solutions might be rigid and shortsighted, or even lead to the heavy burden of manpower. Thus, a more intelligent plan is still urgently required.

\begin{figure}[h]
    \centering
    \includegraphics[width=0.45\textwidth]{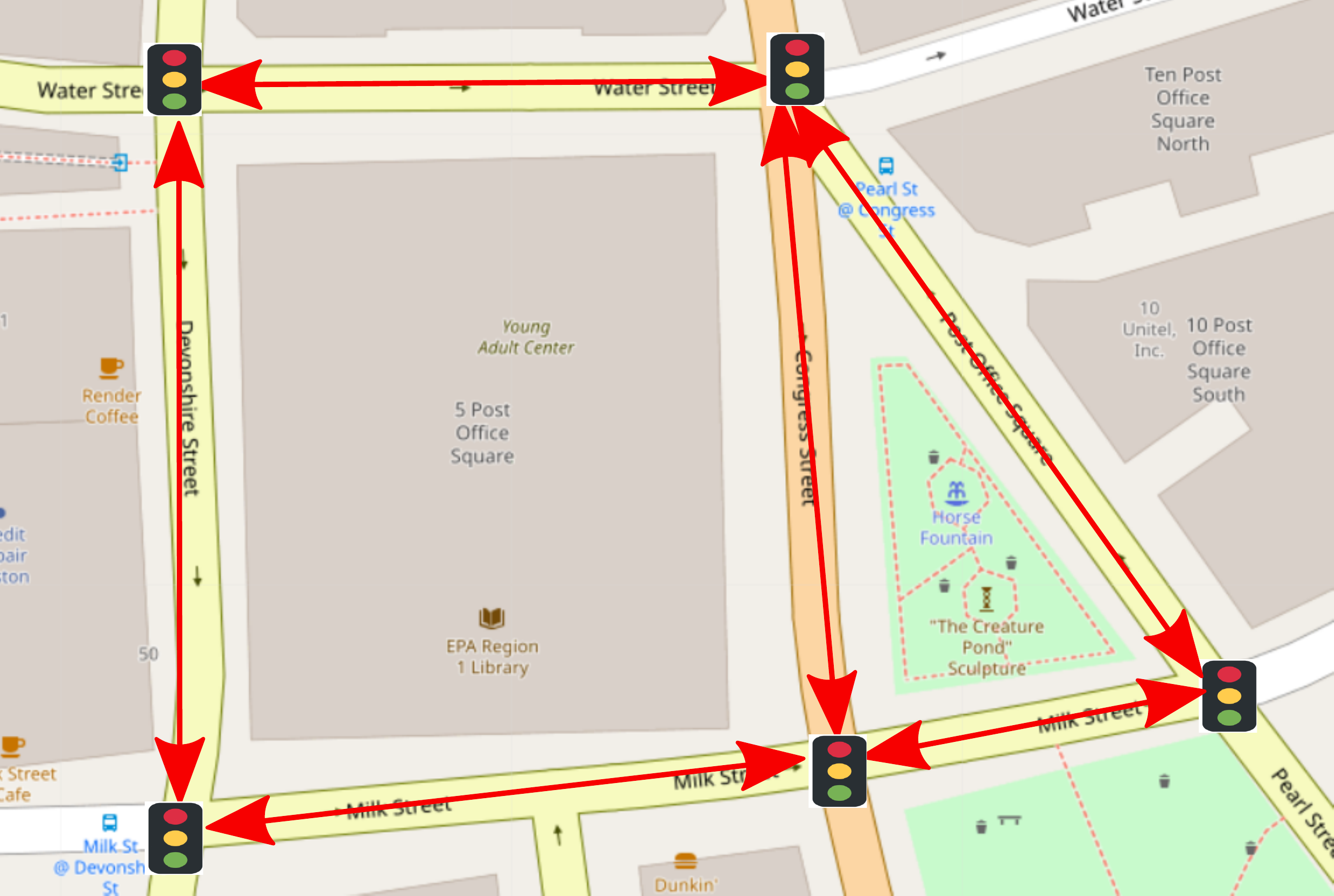}
    \caption{Example of the spatial structure among multiple traffic lights along the road netowrk.}
    \label{fig:intro_1}
\end{figure}

Thanks to the development of data analytic techniques, nowadays, traffic light control has been supported by advanced methods like \emph{reinforcement learning}\textcolor{black}{~\cite{wei2018intellilight, mannion2016experimental,watter2015embed}}, which effectively model the traffic status to make sequential decisions. However, though prior arts performed well, most of them are restricted to the isolated intersections without coordination. Indeed, in a real-world situation, control of traffic light will definitely impact the traffic status and then results in a chain reaction on adjacent intersections. Obviously, mutual influence among multiple intersections should not be neglected during the modeling. To that end, solutions based on \emph{multi-agent} reinforcement learning have been designed~\cite{busoniu2008comprehensive, kuyer2008multiagent}, which further improved the performance. However, they may still face some challenges. First, the dimensionality of action space grows exponentially with the increasing number of agents, which causes dramatic complexity. Second, though distributed models may alleviate the problem of dimension explosion, it is still difficult to formulate the coordination among multiple traffic lights.

Intuitively, when describing the correlation among multiple intersections, we realize that they could be approximately formulated as \emph{graph} structure based on spatial adjacency on the road network as shown in Figure~\ref{fig:intro_1}. The graph structure can be greatly different due to different types of road connections between traffic lights. Similar to information flow in the graph, traffic volume in the current intersection can be naturally split for adjacent intersections, which results in the \textbf{spatial influence} among multiple traffic lights. Thus, when jointly controlling multiple traffic lights to optimize large scale traffic situation, it is critical to model the cooperation structure among multiple traffic lights. Moreover, short periods are spent on traffic flow when moving to adjacent intersections, which further results in the \textbf{temporal dependency} among multiple traffic lights. Therefore, our challenge has been transferred as modeling the spatio-temporal influence among multiple intersections for intelligent traffic light control.

To that end, in this work, we propose a Spatio-Temporal Multi-agent Reinforcement Learning (STMARL) framework for multi-intersection traffic light control. Specifically, we first construct a directional traffic light adjacency graph based on the spatial structure among traffic lights. Then, historical traffic records will be incorporated with current traffic status via Recurrent Neural Network structure. Afterwards, based on the traffic information with temporal dependency, we design a Graph Neural Network based module to model the cooperation structure among multiple traffic lights, which allows the efficient relation reasoning among the traffic lights. Finally, the distributed decision for each traffic light will be made by a deep Q-learning method. Both quantitative and qualitative experiment results have demonstrated the effectiveness of our STMARL framework, which provides an insightful understanding of the influence mechanism among multi-intersection traffic lights. The technical contribution of this paper could be summarized as follows:

\begin{itemize}

\item To the best of our knowledge, we are among the first ones who study the spatio-temporal dependency among multiple traffic lights based on the constructed directional traffic light adjacency graph, with leveraging graph structure for better modeling cooperation mechanism between traffic lights.

\item A novel multi-agent reinforcement learning framework is proposed, in which graph neural network with attention mechanism\textcolor{black}{~\cite{velickovic2017graph}} for iterative relational reasoning and the recurrent neural network is incorporated to model the spatio-temporal dependency.

\item Experiments on both synthetic and real-world datasets validated the effectiveness of our solution compared with several state-of-the-art methods, and further revealed some cooperation mechanism among traffic light agents.
\end{itemize}

\section{Related Works}
\label{sec:relatedwork}
In this section, we briefly review the related works in traffic light control, methodologies of multi-agent reinforcement learning and graph neural networks.

\noindent\textbf{Traffic Light Control.}
 In the literature, traffic light control methods can be mainly divided into three types: predefined fixed-time control~\cite{miller1963settings}, actuated control~\cite{cools2013self} and adaptive traffic control~\cite{mannion2016experimental, el2013multiagent, khamis2014adaptive, abdulhai2003reinforcement}.
The predefined fixed-time control is determined offline using historical traffic data and \textcolor{black}{actuated control is based on current traffic state using predefined rules to decide when to set the green phases (e.g., extending the time of green phase or setting to red phase)}. The main drawback of these two methods is that they do not take into account the long term traffic situation. Therefore, researchers began exploring adaptive traffic light control methods. Following this line, reinforcement learning methods have been used for traffic light control~\cite{kuyer2008multiagent, wei2018intellilight, el2010agent, steingrover2005reinforcement, arel2010reinforcement, el2013multiagent, rizzo2019time} so that the control strategy can be adaptively created based on the current traffic state.

Although reinforcement learning methods have achieved success for traffic light control in one intersection, it's still challenging for the multi-intersection traffic light control task: the curse of dimensionality in the centralized model and coordination problem in a distributed model. Kuyer et al.~\cite{kuyer2008multiagent} \textcolor{black}{and van der Pol et al.~\cite{van2016coordinated}} formulated the explicit coordination among agents using \textcolor{black}{max-plus~\cite{kok2006collaborative}} algorithm which estimates the optimal joint action by sending locally optimized messages among connected agents, ~\textcolor{black}{which needed to handle the combinatorially large joint action space and was computationally expensive. In spite of using \textcolor{black}{max-plus} algorithm, Baker et al.~\cite{bakker2010traffic} handled the partial observability of the traffic state by estimating a belief state using Bayes’rule. However, it did not model the high-order interactions between traffic lights and was computationally expensive. Khamis et al.~\cite{khamis2014adaptive}
extended the framework of~\cite{wiering2000multi} using Bayesian theory and optimize traffic signal control in a multi-objective setting. However, it did not explicitly model the cooperation structure among traffic lights. Chu et al.~\cite{chu2019multi} proposed to utilize independent advantage actor-critic (A2C) instead of Q learning for traffic light control. Although they augmented each agent’s state representation with the state of its neighbors and using a spatial discount factor to adjust the global reward for each agent, they only incorporated the first-order neighbor information for each agent.}
Hua et al.~\cite{presslight19} suggested to use max pressure~\cite{varaiya2013max} as a reward for traffic light control in the arterial network.
Neighbor RL~\cite{arel2010reinforcement} directly concatenated the neighboring intersections' observation into their state representation. However, it did not discriminate neighbors with different traffic situations and only considered the nearest intersection.
Nish et al.~\cite{nishi2018traffic} proposed to used a graph convolution network to extract traffic features of distant roads. Its graph was constructed on lane level where each lane was regarded as a node and vehicle traffic movement connected to two lanes was represented as an edge. The graph size is increasingly larger as the number of intersections becomes larger. It also did not distinguish the traffic flow from different neighbors. Recently, Hua et al.~\cite{colight} introduced a graph attention network for network-level traffic light control. However, 
neighbors of each agent were determined using rules with a predefined and fixed amount.
The traffic flow direction on the graph was also not incorporated.

\textcolor{black}{The above methods control traffic lights based on the physical infrastructure of traffic lights. Instead of using the physical infrastructure of traffic lights, some researchers~\cite{bazzi2016distributed, ferreira2011impact} proposed to control traffic lights with the concept of Virtual Traffic Light (VTL), which is an infrastructure-less traffic control system solely based on
Vehicle-to-Vehicle (V2V) communication. In the concept of VTL, a vehicle in the intersection is selected as a leader, which is responsible for creating and controlling VTL as well as broadcasting traffic signal messages.}

Different from prior arts, in this paper, we proposed STMARL to collectively learn the spatio-temporal dependency among multiple traffic lights based on a constructed intersection-level directional traffic light adjacency graph. \textcolor{black}{Although we focused on controlling physical traffic lights in this paper, we can also apply the proposed method on the virtual traffic lights with a slight modification. Specifically, after selecting the leader in each intersection, we can apply the proposed method on the level of these leaders by regarding these leaders as traffic lights}.


\noindent\textbf{Multi-agent Reinforcement Learning.}
In the setting of multi-agent reinforcement learning~\cite{busoniu2008comprehensive, 
 gupta2017cooperative, panait2005cooperative, foerster2016learning, sukhbaatar2016learning}, agents are optimized to learn cooperative or competitive \textcolor{black}{goals}. \textcolor{black}{Cooperative operation is important in multi-agent systems, such as localization of networked agents and multi-object tracking~\cite{conti2019soft, win2018network, sharma2019decentralized}, robot navigation and autonomous driving~\cite{dunbabin2009experiments, ngai2011multiple}} Independent \textcolor{black}{Deep Q-Networks (DQN)}~\cite{mnih2013playing, tampuu2017multiagent} extends DQN to multi-agent settings where each agent learns its policy independently. Although there is a non-stationary problem for independent DQN, it often works well in practice~\cite{zawadzki2014empirically, tampuu2017multiagent}. To address the issue of reinforcement learning method for multi-agent settings, Lowe et al.~\cite{lowe2017multi} proposed multi-agent actor-critic for mixed cooperative-competitive environments. They adopt the framework for centralized training with decentralized execution for cooperation. 

Note that previous works~\cite{sukhbaatar2016learning, jiang2018graph} mainly design heuristic rules to decide who or how many agents the target agent communicates to, while in this paper, we learn to communicate via the existing spatial structure among agents as well as temporal dependency for multi-intersection traffic light control.

\noindent\textbf{Graph Neural Networks.}
Our proposed method is also related to recent advances of Graph Neural Network (GNN)~\cite{scarselli2009graph, battaglia2018relational, battaglia2016interaction, li2015gated}. GNN has been proposed to learn the structured relationship, which allows for iterative relational reasoning~\cite{zhi2019abductive} through message passing~\cite{gilmer2017neural} on the graph. Battaglia et al.~\cite{battaglia2018relational} introduced a general framework of \emph{Graph Networks} which unified various proposed graph network architectures to support relational reasoning and combinatorial generalization. Recently, some works are trying to explore relational inductive biases in deep reinforcement learning. \textcolor{black}{Wang et al.~\cite{wang2018nervenet} proposed NerveNet for robot locomotion, where it modeled the skeleton of a robot using a discrete graph structure and output actions for different nodes of this robot.} Zambaldi et al.~\cite{zambaldi2018relational} proposed to use relational inductive biases in deep reinforcement learning agents for StarCraft II game. However, there is no explicit graph construction that is learned from the raw visual input. Comparatively, in this paper, we explicitly construct the directional traffic light adjacency graph for modeling the geographical structure information to facilitate the coordination among multi-intersection traffic light control.

\begin{figure}[t]
\begin{tabular}{c c}
\hspace{-4mm}\includegraphics[width=0.2\textwidth]{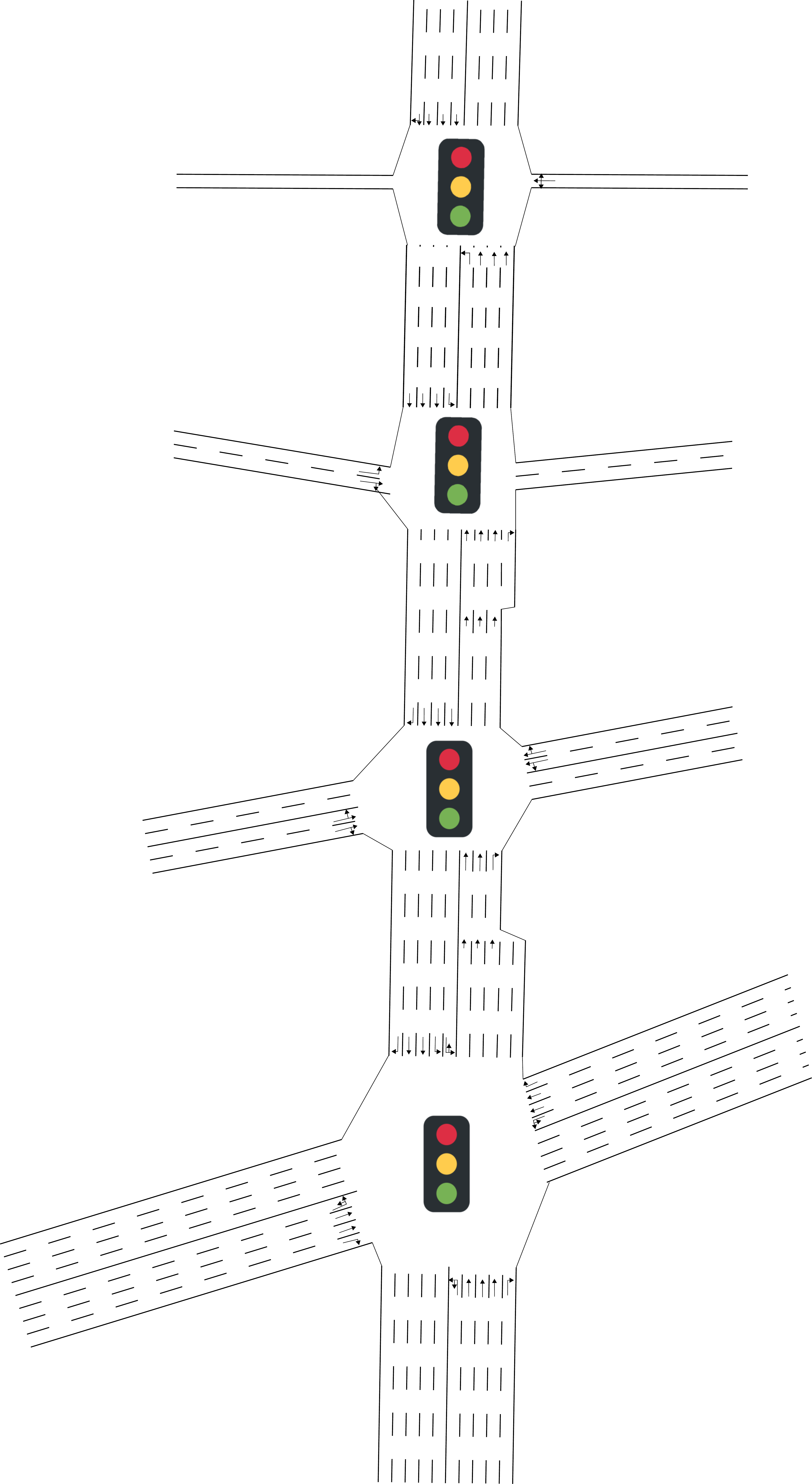} & 
\hspace{-2mm}\includegraphics[width=0.3\textwidth]{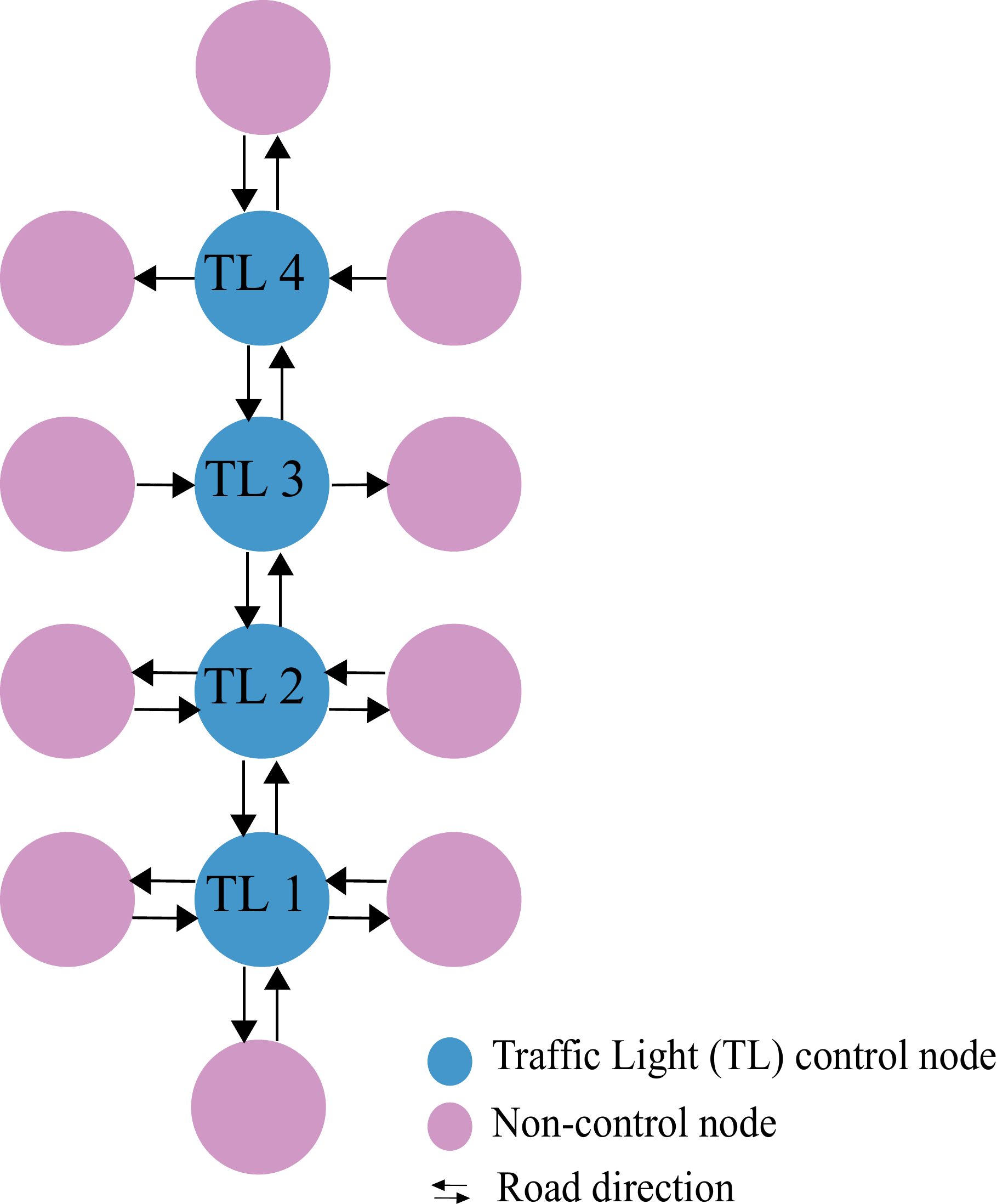}\\
(a) Spatial structure  &   (b)  Constructed graph
\end{tabular}
\caption{An illustration of the traffic light adjacency graph structure in the real world.}
\label{fig:tsl_graph}
\end{figure}

\begin{figure}[t]
    \centering
    \includegraphics[width=0.45\textwidth]{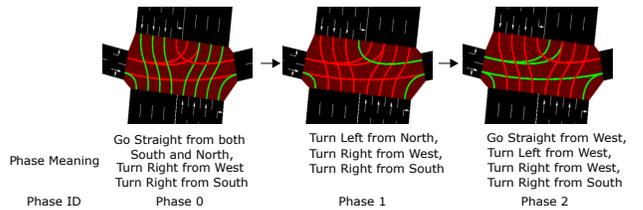}
    \caption{\textcolor{black}{One sample of fixed phase order in a cycle used in the real world with the meaning of each traffic light phase.}}
    \label{fig:phase_order}
\end{figure}
\section{Problem Statement}
\label{sec:problem}
In this section, we will first introduce the construction of the directional \emph{traffic light adjacency graph}, and then formally define our problem.

We first attempt to describe the road network structure in the graph perspective. In a real-world scenario, the structure of the intersections could be complicated. For instance, as shown in Figure~\ref{fig:tsl_graph}~(a), those roads, which are linked to the same intersection, may hold a different \textcolor{black}{number of lanes (e.g., ranged from 1 to 5)} and different \textcolor{black}{passing} constraint (two-way or one way). To describe the complicated settings, we construct the \emph{traffic light adjacency graph} as $G = (V, E)$, as shown in Figure~\ref{fig:tsl_graph} (b). \textcolor{black}{Specifically, $V=\left\{{v}_{i}\right\}_{i=1}^{|V|}$ denotes a set of nodes, where $v_i$ is the $i$-th node's \textcolor{black}{observation information}. The type of nodes includes} the \emph{control nodes} which contain the traffic lights (blue nodes), and \emph{non-control nodes} which indicate the endpoints (pink nodes). \textcolor{black}{The notation of non-control nodes (end-points) is included for the representation of graphical integrity.} At the same time, \textcolor{black}{$E=\left\{\left(e_{k}, {rec}_{k}, {send}_{k}\right)\right\}_{k=1}^{|E|}$ denotes a set of edges, in which $e_k$ indicates the $k$-th edge's \textcolor{black}{observation information}, $rec_k$ is the index of the receiver node and $send_k$ is the index of the sender node of the $k$-the edge.} The $k$-th edge is a directed road connecting two nodes with the type $c(k)$, meaning the number of lanes $l_k$ in this directed road. Definitely, we use a unidirectional edge in $G$ to present each one-way road, and each two-way road is presented by two edges with opposite directions.

Based on the constructed \emph{traffic light adjacency graph}, we then turn to study the problem of multi-intersection traffic light control in the view of \emph{multi-agent reinforcement learning}. Specifically, we treat each traffic light as one \emph{agent}, and the group of traffic light agents is learned cooperatively to maximize the global reward (e.g., minimize the overall queue length in this area). Along this line, the multi-intersection traffic light control problem could be defined as a \emph{Markov Decision Process} (MDP) for $N$ agents within finite steps. Moreover, considering \textcolor{black}{that each traffic light agent receives local noisy observations in the real world~\cite{bakker2010traffic}}, we further extend the MDP problem as \emph{Partially-Observable Markov Decision Process} (POMDP), which can be defined as a tuple $(N, \mathcal{S}, \mathcal{O}, \mathcal{A}, \mathcal{P}, \mathcal{U}, \mathcal{R}, \gamma)$, in which $N$ denotes the number of agents, and the rest are listed as follows:

\begin{figure*}[t]
    \centering
    \includegraphics[width=0.9\textwidth]{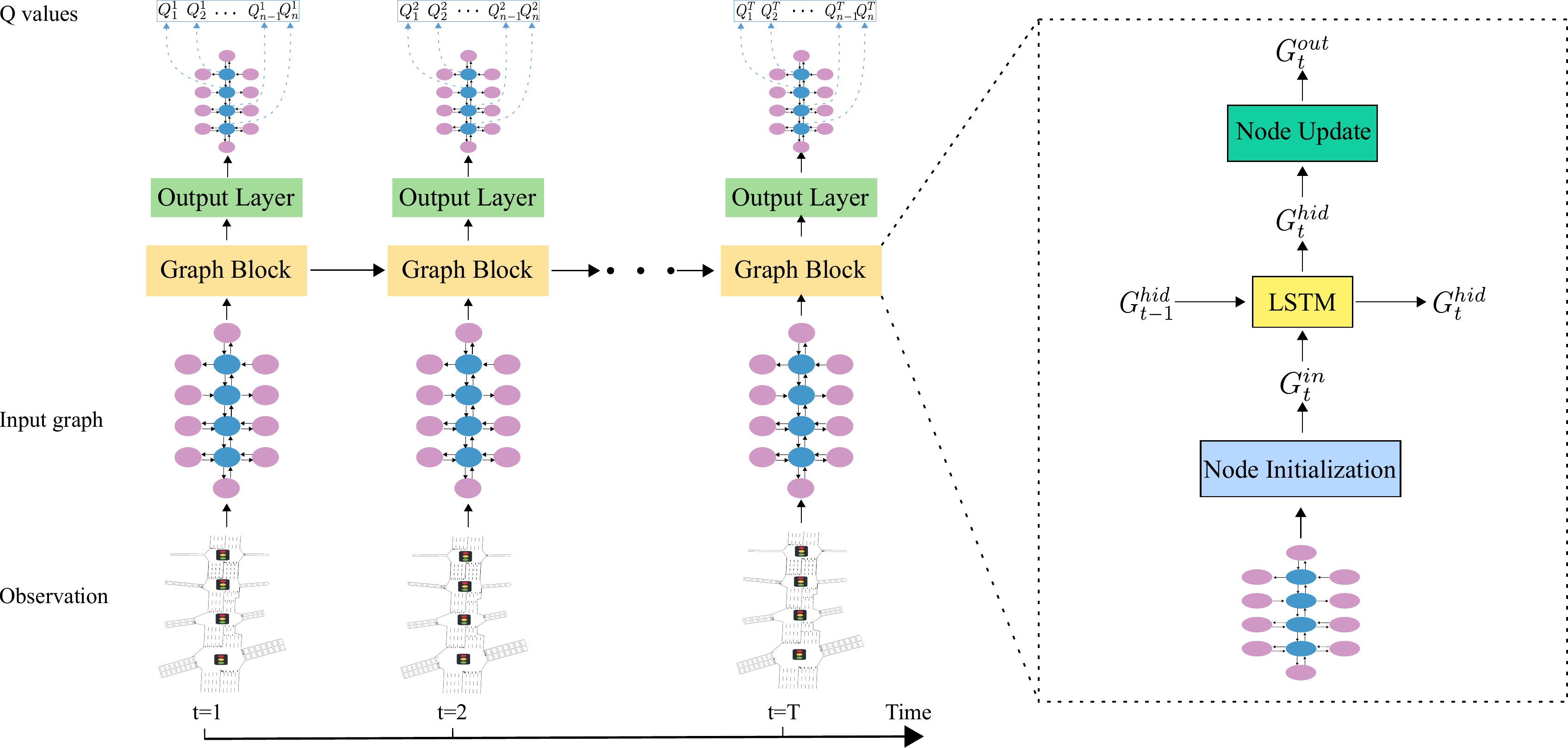}
    \caption{Overall Q-network for Spatio-Temporal Multi-Agent Reinforcement Learning.}
    \label{fig:framework}
\end{figure*}

\begin{table}[]
    \centering
    \caption{Mathematical Notations}\label{tab:notations}
    \small
\begin{tabular}{p{0.4 in}| p{2.7in}}

\hline
Notation    & Description                                            \\
\hline
$o_{i,t}$     & Observation for agent $i$ at time $t$                  \\
$a_{i,t}$     & Action for agent $i$ at time $t$                       \\
$r_{i,t}$     & Reward for agent $i$ at time $t$                       \\
$Q_{i,t}$     & Q value for agent $i$ at time $t$                      \\
$q_l$       & Queue length in the lane $l$                           \\
$n_l$       & Number of vehicles in the lane $l$                     \\
$speed_l$       & Average speed of vehicles on the lane $l$  \\
$e_k$ & \textcolor{black}{Observation information} in the $k$-th edge \\
$v_i$ & \textcolor{black}{Observation information} in the $i$-th node \\
$l_k$       & Lanes number in the $k$-th edge  \\
$c(k)$      & Edge type for the $k$-th edge                          \\
$G$         & $G=(V,E)$ is the constructed traffic light adjacency graph       \\
$G_t^{in}$  & Graph $G$ (node's \textcolor{black}{observation information} is $v_{i,t}^{in}$) after node initialization at time $t$      \\
$G_t^{hid}$ & Output hidden state (node's \textcolor{black}{observation information} is $v_{i,t}^{hid}$) of graph LSTM at time $t$ \\
$G_t^{out}$ & Traffic light graph (node's \textcolor{black}{observation information} is $v_{i,t}^{out}$) after node update at time $t$      \\
$v_{i, t}^{d}$ & Updated node $i$'s \textcolor{black}{observation information} vector after relation reason step $d$ at time $t$      \\
$f_e^{c_k}$ & Edge encoder for edge type of $c_k$                    \\
$\Delta_t$  & Time interval to consider the temporal dependency     \\
$\epsilon$  & Exploration probability                                \\
$\mathcal{D}$  & Memory buffer                                \\\hline
\end{tabular}
\end{table}

\begin{itemize}
\item \textbf{State Space $\mathcal{S}$}: \textcolor{black}{$s_{i,t} \in \mathcal{S}$ is the true system state, which is usually not observable. For agent $i$, the true system state consists of all the complete and accurate information in the traffic
light adjacency graph (i.e., the accurate traffic information in the whole area) at time $t$, which is not directly accessible. Instead, the agent receives the observations}. 
\item \textbf{Observation Space $\mathcal{O}$}: $o_{i,t} \in \mathcal{O}$ \textcolor{black}{is the received observation for agent $i$ in partially observable environment at time $t$.}
In the graph $G$, \textcolor{black}{the observed traffic information on each edge and each node is denoted as $e_k$   and $v_i$:} $e_k = \{\{q_l\}^{l_k}_{l=1}, \{n_l\}^{l_k}_{l=1},$ $\{speed_l\}^{l_k}_{l=1}\}$, $v_i = phaseID_i$ where $q_l, n_l, speed_l$ are the queue length, number of vehicles and average speed of vehicles in lane~$l$ respectively and $phaseID_i$ is the current phase ID for node $i$. \textcolor{black}{Phase ID is the index of a phase in the predefined phase set, where the phase is defined as a combination of valid movements for vehicles. One example of the meaning of phase is illustrated in Figure~\ref{fig:phase_order}.} Note that the partially observability $o_{i,t}$ for agent $i$ is defined as the observed edges information $\{e_k\}_{rec_k=i}$ whose receiver node is node $i$ and the observed node information $v_i$, \textcolor{black}{which represents a local view of each traffic light agent and further motivates the learning of structure dependency among traffic lights. Besides, the observed traffic information is noisy in the realistic environment, e.g., due to the noisy sensors. Therefore, the partial observability in our problem refers to the local noisy observed information for each traffic light agent.}
It is also straightforward to incorporate other complex features into the observation representation, while we focus on designing a novel framework for multi-intersection traffic light control in this paper.

\item \textbf{Action} $\mathcal{A}$: $a_t=\{a_{i,t}\}_{i=1}^{N} \in \mathcal{A}$ is the joint action for all the traffic light agents at time $t$. 
There are mainly two kinds of action settings. One is to determine whether current phase switches to the next phase ~\cite{mannion2016experimental,wei2018intellilight} as shown in Figure~\ref{fig:phase_order} in a fixed phase order, which is due to the constraints and safety issues in the real-world setting. The other is more flexible that action is chosen from the predefined phase set~\cite{colight,wiering2004simulation}. These two settings are tested in the experiment. More experimental details are shown in Section~\ref{sec:simulation_setting}.
\item \textbf{Reward $\mathcal{R}$}: $r_{i,t}$ is the immediate reward for agent $i$ at time $t$. Traffic agent $i$ is optimized to maximize the expected future return $\mathbb { E } \left[ \sum _ { t = 1 } ^ { T } \gamma ^ { t - 1 } r _ { i,t } \right]$, where $\gamma$ is the discount factor. The individual reward $r_{i,t}$ for agent $i$ is $r_{i,t} = -\sum_{l = 1}^{l_i}q_l$,
where $l_i$ is the number of incoming lanes connected to intersection $i$.
\item \textbf{State Transition Probability $\mathcal{P}$}: $p(s_{t+1}|s_{t}, a_t)$ defines the probability of transition from state $s_t$ to $s_{t + 1}$ when all the agents take joint action $a_t$.
\item \textbf{Observation Probability $\mathcal{U}$}: This is the probability of observation $o_ { t + 1 }$, $o _ { t + 1 } \sim \mathcal{U} \left( o _ { t + 1 } | s _ { t + 1 } , a _ { t } \right)$.
\end{itemize}

Based on the formulations above, we can formally define the problem of multi-intersection traffic light control as follows, and related mathematical notations are summarized in Table~\ref{tab:notations}.

\begin{Def}(\emph{Problem Definition}).
Given the \emph{traffic light adjacency graph} $G$, as well as the potential reward $r_{i,t}$ for each action $a_{i,t}$ made by every traffic light agent $i$ at time $t$, we target at making proper decisions $a_{i,t}$ for each traffic light agent $i$, so that the global reward $\sum_{i=1}^N{r_{i,t}}$ will be maximized.
\end{Def}


\section{Spatio-Temporal Multi-Agent Reinforcement Learning}
\label{sec:method}

In this section, we will introduce our Spatio-Temporal Multi-Agent Reinforcement Learning (STMARL) framework in detail for multi-intersection traffic light control.

\subsection{Overview}

The overall framework of STMARL is illustrated in Figure~\ref{fig:framework}. To be specific, we construct the graph consisting of the traffic light agents, and then use the graph block to learn \emph{spatial structure information} on the input graph, with considering the historical traffic state as incorporating \emph{temporal dependency}. 

The module inside the graph block is shown in the right part of Figure~\ref{fig:framework}, which includes:
\begin{itemize}
\item A \emph{node initialization} module, to get the initial node representation;
\item A \emph{recurrent neural network} variant, namely Long Short Term Memory unit (LSTM)~\cite{hochreiter1997long} to summarize historical traffic information in hidden state to learn \emph{temporal dependency};
\item A \emph{node update} module, to update the state of each traffic light.
\end{itemize} 

Along this line, each agent has interactions with other traffic light agents, which is beneficial for the multi-intersection traffic light control at a system level. At the same time, the problem of partial observability will be handled. In the following subsections, we will introduce all these modules in detail. 

\subsection{Base Multi-agent Reinforcement Learning}\label{sec:base_model}
Firstly, we will briefly introduce the base multi-agent reinforcement learning method which is based on the Independent \textcolor{black}{Deep Q-Networks (DQN)} \cite{tan1993multi}, where each agent $i$ learns the independent optimal function $Q_i$ seperately without cooperation. Formally, for each agent $i$, we have the observation $o_i = \{\{q_l, n_l, w_l\}_{l=1}^{l_i}, phaseID_i\}$, where $l_i$ is the number of incoming lanes which are connected to intersection $i$. We target at minimizing the following loss:
\begin{align}
    \mathcal { L } _ { i } \left( \theta _ { i } \right) &= \mathbb { E } _ { (o_{i,t} , a_{i,t} , r_{i,t} , o_ {i, t+1 }) \sim \mathcal{D} } \left[ \left( y_{i,t} - \textcolor{black}{Q_i} \left( o_{i,t} , a_{i,t} ; \theta _ { i } \right) \right) ^ { 2 } \right], \\
    y_{i,t} &= r_{i,t} + \gamma \max _ { a ^ { \prime } } \textcolor{black}{Q_i} \left( o_{i,t+1} , a ^ { \prime } ; \theta _ { i } ^ { tar } \right),
\end{align}
where $\theta _ { i } ^ { tar }$ is the target Q-network updated periodically to stabilize training. Also, $\mathcal{D}$ is the replay buffer to store the state transitions.

For the Independent DQN, to reduce the parameters of the Q-network which scale with the number of traffic light agents, it's reasonable to share the parameters of the Q-network among agents. Specifically, the first encoder layer is separated to handle heterogeneous input information, and parameters for other layers are shared.

\subsection{Learning Spatial Structure Dependency}\label{sec:graph_sec}
Then, we turn to introduce the learning process inside the graph block, which considers the \emph{spatial structure information} between the traffic light agents for better coordination. Generally, traffic light agents in the base model in Section~\ref{sec:base_model} only learn their policy independently without explicit coordination, and observation state $o_{i, t}$ in the base model simply concatenates the traffic information in incoming lanes connected to intersection $i$, which ignores the spatial structure information. Therefore, a more comprehensive framework is required to leverage the spatial structure of these traffic lights for better coordination, so that the global traffic situation will be optimized.

\subsubsection{Node Initialization}
As the traffic information is observed in the graph edges and nodes as stated in Section~\ref{sec:problem}, we firstly introduce the node initialization module.

\noindent\textbf{Edge to Node Update.} In this problem, \textcolor{black}{the observed traffic information  collected by the $k$-th edge is $e_k$}, where $e_k = \{\{q_l\}^{l_k}_{l=1}, \{n_l\}^{l_k}_{l=1},$ $\{speed_l\}^{l_k}_{l=1}\}$.
To preserve the edge direction (e.g., four directions) to the node representation, we use the one-hot representation of the edge observation. \textcolor{black}{For example, suppose there are four incoming edges connected to a traffic light agent, the one-hot representation for these four edges features can be $[e_0, \boldsymbol{0}, \boldsymbol{0}, \boldsymbol{0}], [\boldsymbol{0}, e_1, \boldsymbol{0}, \boldsymbol{0}], [\boldsymbol{0}, \boldsymbol{0}, e_2, \boldsymbol{0}], [\boldsymbol{0}, \boldsymbol{0}, \boldsymbol{0}, e_3]$.} Then, to transform the raw input into embedded observation vector, edge encoder for different edge types is applied per edge to encode the collected message. The observed edge information $e_k$ is updated as follows:
\begin{equation}\label{eq:edge_update}
    e^{\prime}_k = f_e^{c(k)}(e_k).
\end{equation}
To handle the heterogeneous information in the real-world and reduce parameters of edge encoder $f_e^{c(k)}$ for different edge types, separate parameters are used for the first layer of edge encoder to encode the input with 
different dimensions
, but parameters for the other layers are shared. Specifically, we use two-layer MultiLayer Perceptron (MLP) with the \textcolor{black}{Rectified Linear Units (RELU)~\cite{xu2015empirical} activation function.}
After updating the edge information, we aggregate the edges information to the receiver node as follows:
\begin{equation}
    v_{i, t}^{e} = \sum_{k, rec_k = i}  e^{\prime}_k,
\end{equation}
where $rec_k$ is the receiver node of the $k$-th edge.

Then nodes' initial representations are obtained by concatenating $v_{i, t}^{e}$ with the observed node feature $v_{i,t}$ \textcolor{black}{at time $t$}. Here we denote the graph with initial node values as $G_t^{in}$ with node \textcolor{black}{observation information} $V = \{v_{i, t}^{in}\}_{i=1}^{|V|}$, where initial representation $v_{i,t}^{in}$ at time $t$ is obtained as follows:
\begin{equation}
    v_{i, t}^{in} = f_v(v_{i, t}^{e} \| v_{i, t}),
\end{equation}
\textcolor{black}{where $\|$ represents the concatenation operation} and $f_v$ is one-layer MLP. For $phaseID$ in node feature $v_i$, we use one-hot representation.

\subsubsection{Node Update}
\label{sec:node_update}
Then, we turn to introduce the node update module to model the interaction relationship among these agents. Here,  we use attention mechanism~\cite{vaswani2017attention, velickovic2017graph} to leverage the spatial structure information and perform relation reasoning among these agents. Specifically, at relation reasoning step $d$, the input node vector consists of both the initial node vector $v_{i, t}^{in}$ and the node vector $v^{d-1}_{i, t}$ in previous relation reasoning step $d-1$:
\begin{equation}
    \hat{v}_i = [v_{i, t}^{in} \| v^{d-1}_{i, t}],
\end{equation}
\textcolor{black}{where $\|$ represents the concatenation operation.}

Then, we compute the attention score $\alpha_{ij}$ between node $i$ and its sender nodes \textcolor{black}{$j \in \{ send_k\}_{rec_k=i,k=1:|E|}$}, 
\begin{equation}\label{eq:node_att}
    \alpha _ { i j } = \frac { \exp \left( f \left( w_a^{T} \left[ \hat{v} _ { i } \| \hat{v} _ { j } \right] \right) \right) } { \sum _ { k \in \mathcal { N } _ { i } } \exp \left( f \left( w_a  ^ { T } \left[ \hat{v} _ { i } \| \hat{v} _ { k } \right] \right) \right) },
\end{equation}
\textcolor{black}{where $\|$ represents the concatenation operation.} $w_a$ is the trainable attention weight vector and $f$ is the nonlinear activation function and here we use \textcolor{black}{Exponential Linear Unit(ELU)}~\cite{clevert2015fast} function.
Then the aggregated attention information $\overline{v}_i$ from the sender nodes for node $i$ is:
\begin{equation}\label{eq:node_agg}
    \overline{v}_i = \sum_{\textcolor{black}{j \in \{ send_k\}_{rec_k=i,k=1:|E|}}} \alpha _ { i j } \hat{v}_j.
\end{equation}
Finally, the node vector $v_{i, t}^{d}$ is updated based on its own information and the aggregated neighbor information:
\begin{equation}\label{eq:node_update}
    v_{i, t}^{d} = g([ \hat{v} _ { i } \| \overline{v}_i ]),
\end{equation}
\textcolor{black}{where $\|$ represents the concatenation operation} and $g$ is one-layer MLP with RELU activation in the MLP output layer.

The above node update process is for one step relation reasoning. Multi-step relation reasoning can be performed to capture the high-order interaction among the agents. \textcolor{black}{For example, if relation reasoning step is 2, the traffic light agent can aggregate information from its first-order and second-order neighbors.} The output graph at time $t$ is denoted as $G^{out}_t$ with $V = \{v_{i, t}^{d}\}_{i=1}^N$.

\subsection{Learning Temporal Dependency}\label{sec:temporal_d}
Moreover, we attempt to learn \emph{temporal dependency} to incorporate the historical traffic state information. As the traffic state is highly dynamic over the time, to model the temporal dependency and handle the partial observability in the POMDP problem, we use the recurrent neural network to incorporate the historical traffic information. \textcolor{black}{Using recurrent neural network to summarize the history of observations is one approach to handle the partial observability in POMDP~\cite{hausknecht2015deep, foerster2016learning,heess2015memory}.} Specifically, we process the nodes in current input traffic state graph $G^{in}_t$ and the last time hidden graph $G^{hid}_{t-1}$ using \textcolor{black}{Long Short Term Memory unit (LSTM)~\cite{hochreiter1997long}}. The output hidden graph $G_{t}^{hid}$ with nodes $V = \{v_{i, t}^{hid}\}_{i=1}^N$. \textcolor{black}{The hidden state $v_{i, t}^{hid}$ is computed as follows using LSTM} as follows:
\begin{align*}
    i _ { t } & = \sigma \left( W _ { i } \left[ { v} _ { i,t }^{in} , v _ { i, t - 1 } ^ { hid } \right] + b _ { i } \right) , \\ f _ { t } &= \sigma \left( W _ { f } \left[ { v} _ { i,t }^{in} , v _ { i, t - 1 } ^ { hid } \right] + b _ { f } \right), \\ \widetilde { C } _ { t } & = \tanh \left( W _ { C } \left[ { v} _ { i,t }^{in} , v _ { i, t - 1 } ^ { hid } \right] + b _ { C } \right) , \\
    C _ { t } &= f _ { t } \odot C _ { t - 1 } + i _ { t } \odot \widetilde { C } _ { t }, \\ 
    o _ { t } & = \sigma \left( W _ { o } \left[ { v} _ { i,t }^{in} , v _ { i, t - 1 } ^ { hid } \right] + b _ { o } \right) , \\
    v_{i, t}^{hid} &= o _ { t } \odot \tanh \left( C _ { t } \right) ,
\end{align*}

where, $W _ { f } , W _ { i } , W _ { C } , W _ { o } , b _ { f } , b _ { i } , b _ { C } , b _ { o }$ are parmeters of weight matrices and biases. $\odot$ represents element-wise multiplication and $\sigma$ is the sigmoid function. The above update process is denoted in short as:
\begin{equation}\label{eq:gate_graph}
    v_{i, t}^{hid} = LSTM(v_{i,t}^{in}, v_{i, t - 1}^{hid}).
\end{equation}

After getting the gated hidden graph $G_{t}^{hid}$ at time step $t$, \emph{Node update} process in~\ref{sec:node_update} is performed on the updated traffic input graph $G_{t}^{hid}$ as is shown in the right part of Figure~\ref{fig:framework}.
    
\renewcommand{\algorithmicrequire}{\textbf{Input:}}
\renewcommand{\algorithmicensure}{\textbf{Output:}}
\begin{algorithm}[t]
\caption{\mbox{Spatio-Temporal Multi-Agent Training}}
\centering
\label{alg:model_training}
\begin{algorithmic}[1]
\REQUIRE Traffic light adjacency graph G = (V, E).
\ENSURE Q-network with parameter $\theta$.
\STATE Initialize the parameters of Q-network and target Q-network.
\FOR {epoch = 1 to max-epochs}
\STATE Reset the environment.
\FOR {$t = 0$ to $T$}
\STATE Get observation $o_t$ of the traffic light adjacency graph.
\FOR {agent $i = 1$ to $N$}
\STATE Compute $v_{i,t}^{hid}$ using Eq(\ref{eq:gate_graph}). 
\COMMENT{learning temporal dependency}
\STATE Compute node update result $v_{i, t}^{out}$ using Eq(\ref{eq:node_update}).  \hfill \COMMENT{learning spatial structure dependency}
\STATE Compute Q values $Q_{i,t}$ using Eq(\ref{eq:output_q}).
\STATE With probability $\epsilon$ pick random action $a_{i,t}$, else $a_{i,t} = max_{a^{\prime}} Q_{i,t}(a^{\prime})$.
\ENDFOR
\STATE Execute joint action $a_t = \{a_{i,t}\}_{i=1}^N$ in the environment and get reward $r_t = \{r_{i,t}\}_{i=1}^N$ and next observation $o_{t+1}$ .
\STATE Store transition $(o_t, a_t, o_{t+1}, r_t)$ into $\mathcal{D}$.
\ENDFOR
\FOR {$c=1$ to $C1$}
\STATE Sample a random batch of transitions over continues time interval $\Delta_t$: $\{o_t , a_t , r_t , o _ { t+1 }\}_{t=1}^{\Delta t}$ from $\mathcal{D}$.
\STATE For each agent $i$ at time $t$ compute target: $$y_{i,t} = r_{i,t} + \gamma \max _ { a ^ { \prime } } Q_{i, t + 1} \left(  a ^ { \prime } ; \theta ^ { tar } \right).$$
\STATE Update Q-network: $$\theta \leftarrow \theta - \nabla_{\theta}{\sum_{i=1}^N\sum_{t=1}^{\Delta_t}( y_{i, t} - Q_{i,t} ( a_{i,t} ; \theta)) ^ { 2 }}.$$
\ENDFOR
\ENDFOR
\end{algorithmic}
\end{algorithm}

\subsection{Output Layer}
Finally, the distributed decision for each traffic light agent is now available with the learned representations for traffic lights. Based on the above learned \emph{spatial-temporal dependency} representation, at each time $t$, the decision is made for each traffic light agent. As there are two kinds of nodes: traffic light control nodes and non-control nodes shown in Figure~\ref{fig:tsl_graph}, we process for the traffic light control nodes. For these traffic light agents, we use residual connection which concatenate the initial node feature vector $v_{i, t}^{in}$ and the updated node feature vector $v_{i, t}^{out}$:
\begin{equation}
    x_{i, t} = [v_{i, t}^{in} \| v_{i, t}^{out}],
\end{equation}
\textcolor{black}{where $\|$ represents the concatenation operation.}

Then, Q value for each agent $i$ at time $t$ is computed as follows:
\begin{equation}\label{eq:output_q}
    Q_{i, t} = \phi(x_{i, t}),
\end{equation}
where $\phi$ is two layer MLP with RELU activation, $Q_{i, t} \in \mathbb{R}^{|\mathcal{A}|}$ and we denote $Q_{i, t}(a)$ as the output Q value for action $a$.

\subsection{Training}
Besides, we briefly introduce the technical solution of training process. During training, we store the observations into the replay buffer $\mathcal{D}$ for experience replay~\cite{mnih2015human}. As the observations are in the edges of graph $G$, we denote the observation at time $t$ as $o_t = \{\{e_{k,t}\}_{k=1}^{|E|}, \{v_{i,t}\}_{i=1}^{N}\}$. We store the transition $(o_t, a_t, o_{t+1}, r_t)$ into $\mathcal{D}$, where joint action $a_t = \{a_{i,t}\}_{i=1}^N$ and reward for each agent $r_t = \{r_{i,t}\}_{i=1}^N$. The training loss of Q-network for STMARL model is:
\begin{align}
    \mathcal { L } ( \theta ) &= \mathbb { E } _ { \{o_t , a_t , r_t , o _ { t+1 }\}_{t=1}^{\Delta t} \sim \mathcal{D} }\sum_{i=1}^N\sum_{t=1}^{\Delta_t}( y_{i, t} - Q_{i,t} (a_{i,t} ; \theta)) ^ { 2 } , \label{eq: loss_q} \\ 
    y_{i,t} &= r_{i,t} + \gamma \max _ { a ^ { \prime } } Q_{i, t + 1} \left(a ^ { \prime } ; \theta ^ { tar } \right),
\end{align}
where $\Delta t$ is the time interval. 
We use recurrent neural network over the continuous time interval $\Delta t$ to learn temporal dependency
as stated in ~\ref{sec:temporal_d}. The influence of the temporal dependency interval is illustrated in experiment part. To stabilize training, we update the model at the end of each episode. Detailed training algorithm for Spatio-Temporal Multi-Agent Training is listed in Algorithm~\ref{alg:model_training}.

\subsection{Time Complexity Analysis}\label{Sec:time_complexity}
Here we analyze the time complexity of the STMARL model to show the scalability by learning spatial-temporal dependency. The following two assumptions are made that the \emph{node initialization} can be performed concurrently for each node and for one node, the interaction with its neighbor nodes can perform separately. The neural network hidden layer size is assumed as $h$. Then based on STMARL model structure, the time complexity is computed as follows:
\begin{itemize}
    \item For \emph{node initialization}, the complexity is $d_kh + h^2 + h^2$, where $d_k$ is the edge input feature size and the small size of $phaseID$ in node feature is ignored;
    \item For \emph{node update}, the complexity to learn temporal dependency over time interval $\Delta_t$ using LSTM is $4h(h+h)\Delta_t$. For one step node update, time complexity is $4h \times 4h + 4h \times h$. Thus, time complexity for $L$ step relation reasoning is $(8\Delta_t + 20)h^2L$;
    \item For the \emph{output layer}, time complexity is $2h^2 + h^2$.
\end{itemize}
Therefore, the overall time complexity is \textcolor{black}{$O(d_kh + 5h^2 + (8\Delta_t + 20)h^2L) \approx O(\Delta_th^2L)$}, which scales linearly with the temporal dependency interval $\Delta_t$ and is irrelevant to the number of intersections.

\begin{table}[]
\centering
\caption{Statistics of the real-world traffic datasets.}\label{table:dataset}
\begin{tabular}{c|ccccc}
\hline
\multirow{2}{*}{Dataset}        & \multirow{2}{*}{Time Range} & \multicolumn{4}{c}{Arrival Rate (vehicles/300s)}                                                     \\ \cline{3-6} 
                                &                             & Mean                    & Std                    & Min                     & Max                     \\ \hline
\multirow{2}{*}{$D_{Hangzhou}$} & \multirow{2}{*}{-}          & \multirow{2}{*}{544.83} & \multirow{2}{*}{99.19} & \multirow{2}{*}{375.00} & \multirow{2}{*}{668.00} \\
                                &                             &                         &                        &                         &                         \\
\multirow{2}{*}{$D_{Hefei}$}    & 11/06/2018-                 & \multirow{2}{*}{443.95} & \multirow{2}{*}{38.78} & \multirow{2}{*}{368.00} & \multirow{2}{*}{551.00} \\
                                & 11/12/2018                  &                         &                        &                         &                         \\ \hline
\end{tabular}
\end{table}

\begin{table}[]
\fontsize{8}{10}\selectfont
\centering
\caption{Traffic light phase configurations for different traffic lights in the real-world dataset $D_{Hefei}$.}\label{table:phase_config}
\begin{tabular}{c|c}
\hline
Traffic Light ID & Phase ID      \\ \hline
Traffic Light 1  & 0, 1, 2, 3, 4 \\
Traffic Light 2  & 5, 6, 7, 8    \\
Traffic Light 3  & 9, 10, 11     \\
Traffic Light 4  & 12, 13, 14    \\ \hline
\end{tabular}
\end{table}

\section{Experiments}
\label{sec:exp}
In this section, we conduct both quantitative and qualitative experiments to validate the effectiveness of the proposed STMARL model for multi-intersection traffic light control.

\subsection{Experimental Setup}
\subsubsection{Synthetic Dataset}
In the experiment, synthetic data is generated to test our model under various flexible traffic patterns. We generate these datasets after analyzing the real-world traffic flow data. The details are introduced as follows:
\begin{itemize}
    
    \item $Unidirec_{6 \times 6}$: A $6 \times 6$ grid network with unidirectional traffic from West to East and South to North. The traffic flow is generated using Bernoulli distribution with probability 0.2 and the maximum number of arrival vehicles is limited to 3 in every second for stable simulation.
    
    \item $Bidirect_{6 \times 6}$: A $6 \times 6$ grid network with bidirectional traffic in both West-East and South-North direction. This traffic flow is generated using Bernoulli distribution with probability 0.1 and we stabilize the simulation by setting the maximum number of arrival vehicles as 4 in every second.
    
\end{itemize}

\subsubsection{Real-world Datasets}
Two real-world datasets are used here:
\begin{itemize}
\item $D_{Hangzhou}$~\footnote{\url{https://github.com/wingsweihua/colight/tree/master/data/Hangzhou/4_4/anon_4_4_hangzhou_real_5734.json}}: A publicly available dataset for city Hangzhou, China. The road structure in this dataset is a $4 \times 4$ grid and the duration of the traffic flow is one hour.
\item $D_{Hefei}$: This real-word dataset is collected from Hefei, China, which consists of four heterogeneous intersections as shown in Figure 1 (a). The information of vehicles and roads are recorded by the camera in the nearby intersection facing the vehicles, as well as corresponding timestamps over the time period from 11/06/2018 to 11/12/2018. After analyzing these records, the trajectory for each vehicle could be captured. As shown in Table~\ref{table:dataset}, the traffic arrival rate is significantly variant. For performance comparison in $D_{Hefei}$, we use the traffic flow during the most peak hour in one day. Further
analysis of the performance of the whole day will be introduced in Section~\ref{sec:qualitative}.
\end{itemize}


\subsubsection{Simulation Setting} \label{sec:simulation_setting}
For synthetic datasets and $D_{Hangzhou}$, we used four phases to control the traffic movements for the intersection, i.e., WE-Straight (Going Straight from both West and East), WE-Left (Turning Left from both West and East), SN-Straight (Going Straight from both South and North), SN-Left (Turning Left from both South and North). The action of each traffic light is chosen from these four phases.

For dataset $D_{Hefei}$, we adopt the traffic phases currently applied in the real world during that time period as shown in Table~\ref{table:phase_config} (detailed description of each phase meaning is attached in the appendix). It can be seen that there are different kinds of traffic phases for each intersection. For simplicity, we adopt the switch setting for action selection as shown in Section~\ref{sec:problem}. Therefore, for each agent $i$, $a_{i,t} \in \{0, 1\}$ indicating switch to the next phase (1) or keep current phase (0). Besides, every phase change is followed by a 3-second yellow light. \textcolor{black}{One example of the 
executed Phase ID sequence of Traffic Light 1 (phase order is 0,1,2,3,4 as shown in Table ~\ref{table:phase_config}) can be $0 \rightarrow 0 \rightarrow 1 \rightarrow 2 \rightarrow 2 \rightarrow 2 \rightarrow 3 \rightarrow 3 \rightarrow 4$.}

To simulate the traffic status under different traffic settings, we then utilized a traffic simulator called CityFlow~\footnote{http://cityflow-project.github.io}~\cite{zhang2019cityflow, presslight19} for large scale traffic network simulation. Both synthetic and real datasets are fed into the simulator for simulation.

\begin{table}[]
\fontsize{8}{10}\selectfont
\centering
\caption{Parameter settings of our model.}\label{table:parameter}
\begin{tabular}{c|c}
\hline
Parameters         & Values                                                                       \\ \hline
Memory buffer size  & 50 episodes                                                                       \\
Model update step  & 1 episode                                                                    \\
Target update step $C$ & 2 episodes                                                                    \\
$C1$           & 3000                                                                         \\
$\gamma$           & 0.99                                                                         \\
$\epsilon$         & 1 $\rightarrow$ 0.05(linear decay) \\
Relation reasoning step $d$           & 2                                                                        \\
batch size         & 16                                                                           \\ \hline
\end{tabular}
\end{table}

\subsubsection{Evaluation Metric}
Following the previous researches~\cite{wei2018intellilight, mannion2016experimental, colight}, we adopt the commonly used \textbf{average travel time} metric to evaluate the performance of different methods. This metric is defined as the average travel time for all the vehicles traveling from their origins to destinations, which people care the most in practice. This average travel time $avgt$ is calculated as:
\begin{equation*}
    avgt = \frac{1}{N_c}\sum_{i=1}^{N_c}t_{i,ed} - t_{i, st},
\end{equation*}
where $N_c$ is the toal number of cars entering this area. $t_{i, st}$ and $t_{i, ed}$ are the arrival and depart time for the $i$-th car.
\begin{table}[t]
\fontsize{8}{10}\selectfont
\centering
\caption{Summary of the learning-based comparison methods.}\label{table:model_cmpare}
\begin{tabular}{c|ccc}
\hline
Methods     & \begin{tabular}[c]{@{}c@{}}Traffic Light \\ Adjacency Graph\\  (intersection-level)\end{tabular} & \begin{tabular}[c]{@{}c@{}}Neighbor \\ Attention\end{tabular} & \begin{tabular}[c]{@{}c@{}}Temporal \\ Dependency\end{tabular} \\ \hline
Max-Plus~\cite{kuyer2008multiagent}    & Undirected                                                                                       & $\times$                                                      & $\times$                                                       \\
Neighbor RL~\cite{arel2010reinforcement} & Undirected                                                                                         & $\times$                                                      & $\times$                                                       \\
GCN-lane~\cite{nishi2018traffic}         & $\times$                                                                                         & $\times$                                                      & $\times$                                                       \\
GCN-inter         & Undirected                                                                                         & $\times$                                                      & $\times$                                                       \\
Colight~\cite{colight}     & $\times$                                                                                         & $\checkmark$                                                  & $\times$                                                       \\ \hline
STMARL-ST        & $\times$                                                                                         & $\times$                                                      & $\times$                                                       \\
STMARL-T  & $\checkmark$                                                                                     & $\checkmark$                                                  & $\times$                                                       \\
STMARL-S  & $\checkmark$                                                                                     & $\times$                                                  & $\checkmark$                                                       \\
STMARL      & $\checkmark$                                                                                     & $\checkmark$                                                  & $\checkmark$                                                   \\ \hline
\end{tabular}
\end{table}
\subsubsection{Implementation Details}
The parameters for our model are summarized in Table~\ref{table:parameter}. Particularly, each action of agent will last for 10 seconds to avoid frequent phase switch. Besides, the temporal dependency interval $\Delta t$ was searched among $\{3, 5, 10, 15, 20\}$, the $\epsilon$ for $\epsilon$-greedy policy was linearly decayed for the first 10 episodes, and the hidden size for both edge encoder, output layer and LSTM were set to 64.
\textcolor{black}{The activation function was searched among $\{\rm RELU, \rm ELU, \rm tanh\}$ and the number of MLP layers was searched among $\{1, 2\}$.} Finally, all the parameters were initialized using He initialization in~\cite{he2015delving}, and then trained using Adam~\cite{kingma2014adam} algorithm with learning rate as 0.001 and gradient clipping value as 10.

\begin{table*}[t]
\fontsize{8}{10}\selectfont
\centering
\caption{Performance comparison on the synthetic data and real-world data w.r.t. average travel time (in seconds, the lower the better). '*' indicates the improvement of STMARL over the best baseline is significant based on paired t-test at the significance level of $p < 0.01$. Result on $D_{Hefei}$ is averaged over seven days.}\label{table:main_result}
\begin{tabular}{c|cccc}
\hline
Methods  & $Unidirec_{6 \times 6}$                   & $Bidirect_{6 \times 6}$             & $D_{Hangzhou}$       & $D_{Hefei}$                   \\ \hline
 Fixed-time &608.34   &481.35   &572.15   & 153.63   \\
 MaxPressure &289.04  &227.95  &475.89 &92.12  \\\hline
 Max-Plus & 891.68  & 1127.19  & 1123.55  & 111.56  \\ 
 Neighbor RL &213.98  &182.20  &331.95  &80.45  \\
 GCN-lane&611.28   &227.31  &690.36  &130.64  \\
 GCN-inter&276.85  &213.36  &401.25  &98.41  \\
 Colight&273.92  &194.84  &394.10  &249.70  \\\hline
 \textbf{STMARL}&$\textbf{205.34}^*$  &$\textbf{180.31}^*$ &$\textbf{319.14}^*$  &$\textbf{63.86}^*$  \\ \hline
\end{tabular}
\end{table*}
\subsubsection{Compared Methods}
To validate the effectiveness of STMARL framework, several state-of-the-art methods are selected as baseline methods. There are mainly two categories:
transportation methods and reinforcement learning methods.

\begin{table}[]
\fontsize{8}{10}\selectfont
\centering
\caption{Ablation study of the model component. '*' indicates the improvement of STMARL over each ablation model is significant based on paired t test with $p < 0.01$.}\label{table:ablation_study}
\scalebox{0.9}{\begin{tabular}{c|cccc}
\hline
Methods                 & $Unidirec_{6 \times 6}$ & $Bidirect_{6 \times 6}$ & $D_{Hangzhou}$ & $D_{Hefei}$ \\\hline
STMARL-ST                    &229.75                         &194.42                         &329.97                &71.94             \\
STMARL-T              &214.68                         &180.91                         &322.17                &66.02             \\
STMARL-S              &218.09                         &181.13                         &329.64                &65.58             \\
 \textbf{STMARL} &$\textbf{205.34}^*$                         &$\textbf{180.31}^*$                         &$\textbf{319.14}^*$                &$\textbf{63.86}^*$             \\ \hline
\end{tabular}}
\end{table}
\begin{figure*}
    \centering
    \subfigure[$Unidirec_{6 \times 6}$]{\includegraphics[width=0.4\textwidth]{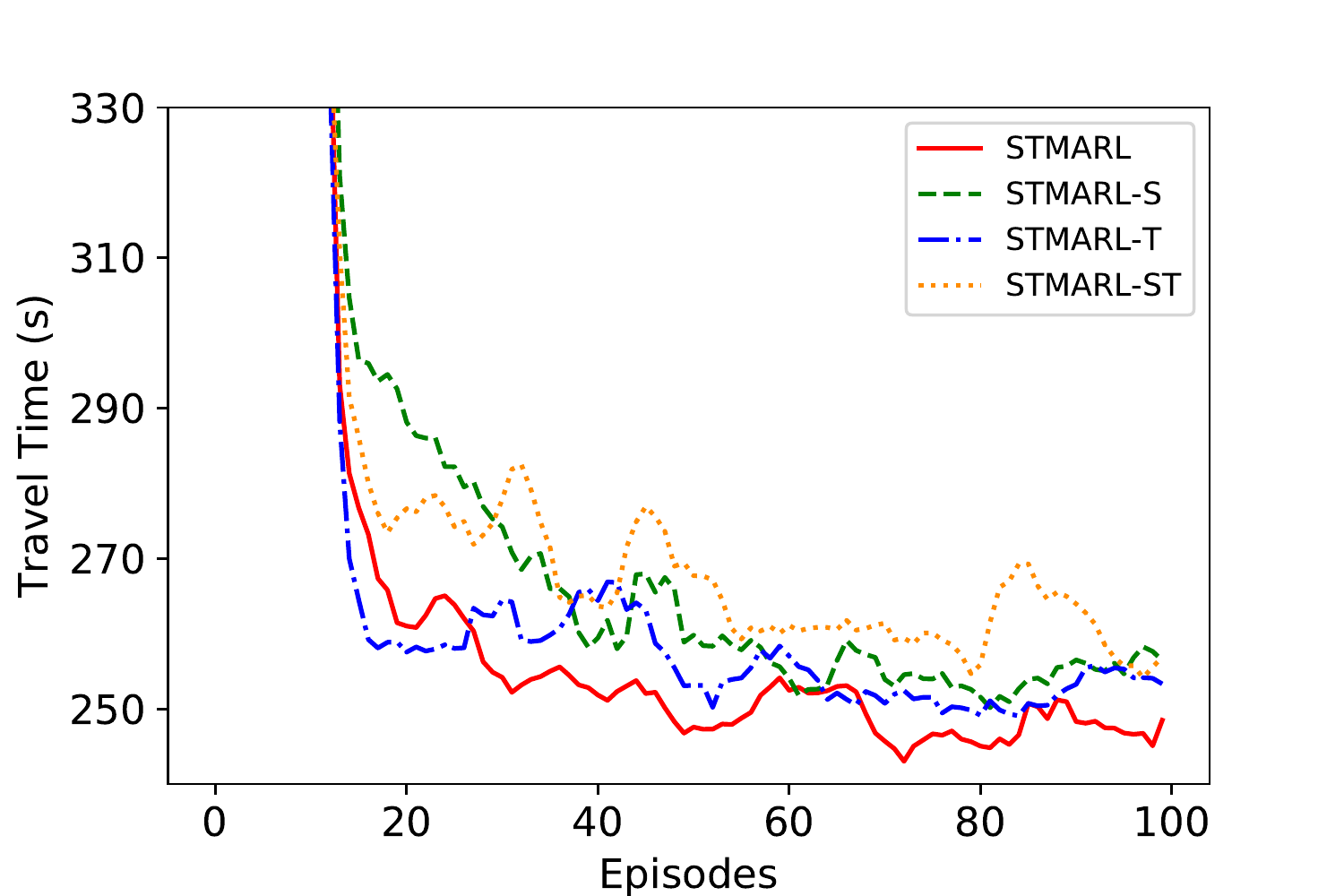}}\label{fig:train_curve_6_6_uni}
    \subfigure[$Bidirect_{6 \times 6}$]{\includegraphics[width=0.4\textwidth]{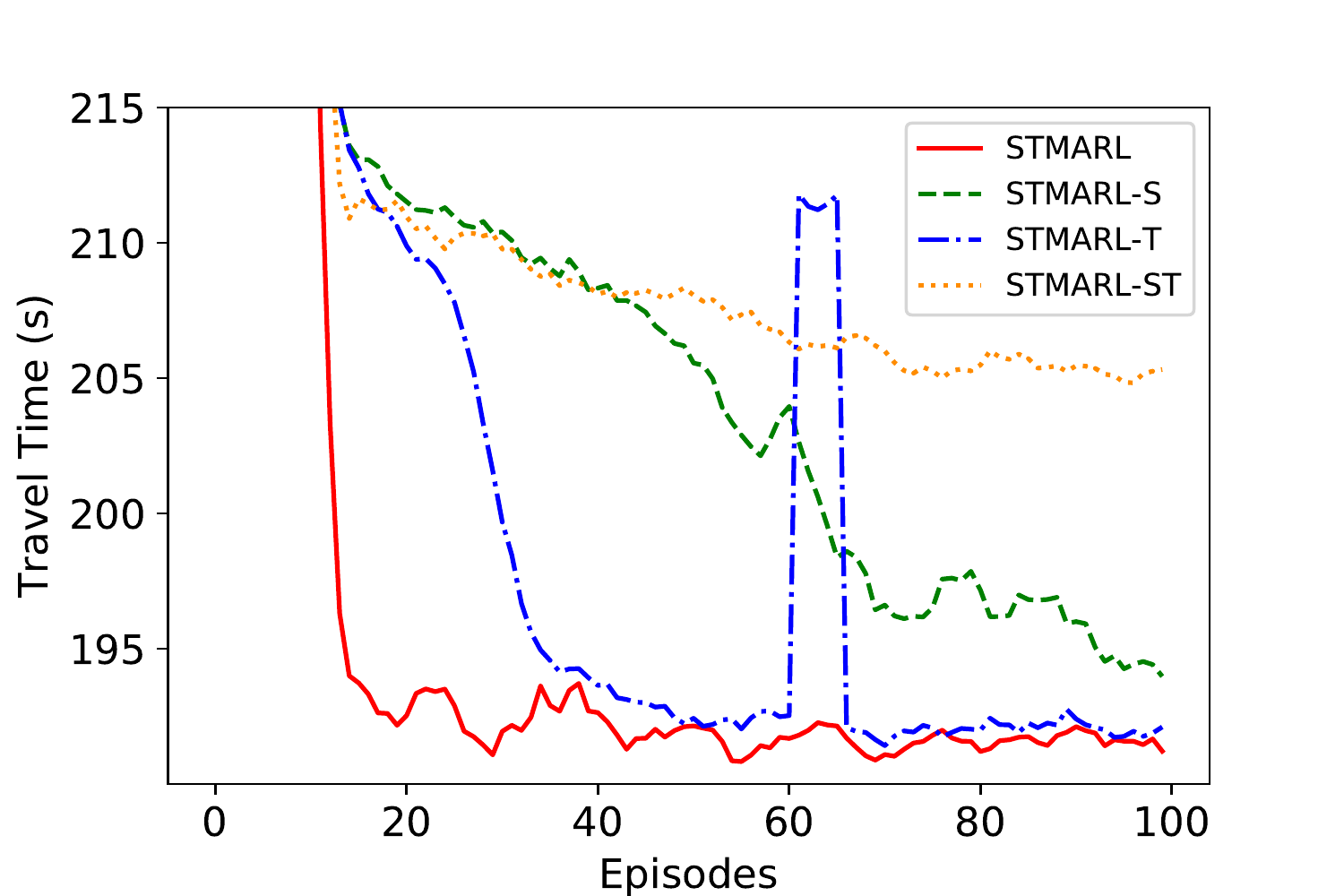}}\label{fig:train_curve_6_6_uni}
    \subfigure[$D_{Hangzhou}$]{\includegraphics[width=0.4\textwidth]{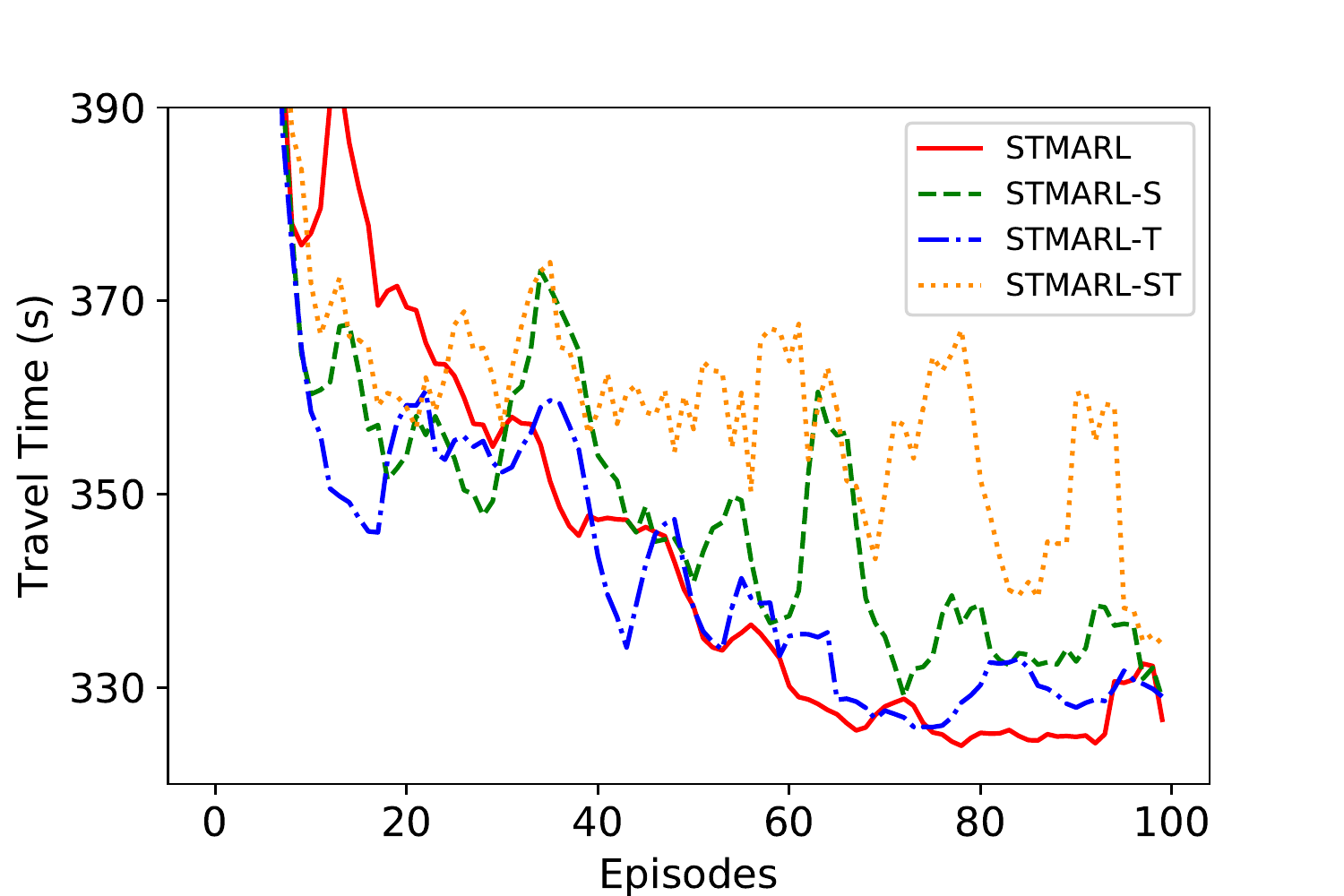}}\label{fig:train_curve_hz}
    \subfigure[$D_{Hefei}$]{\includegraphics[width=0.4\textwidth]{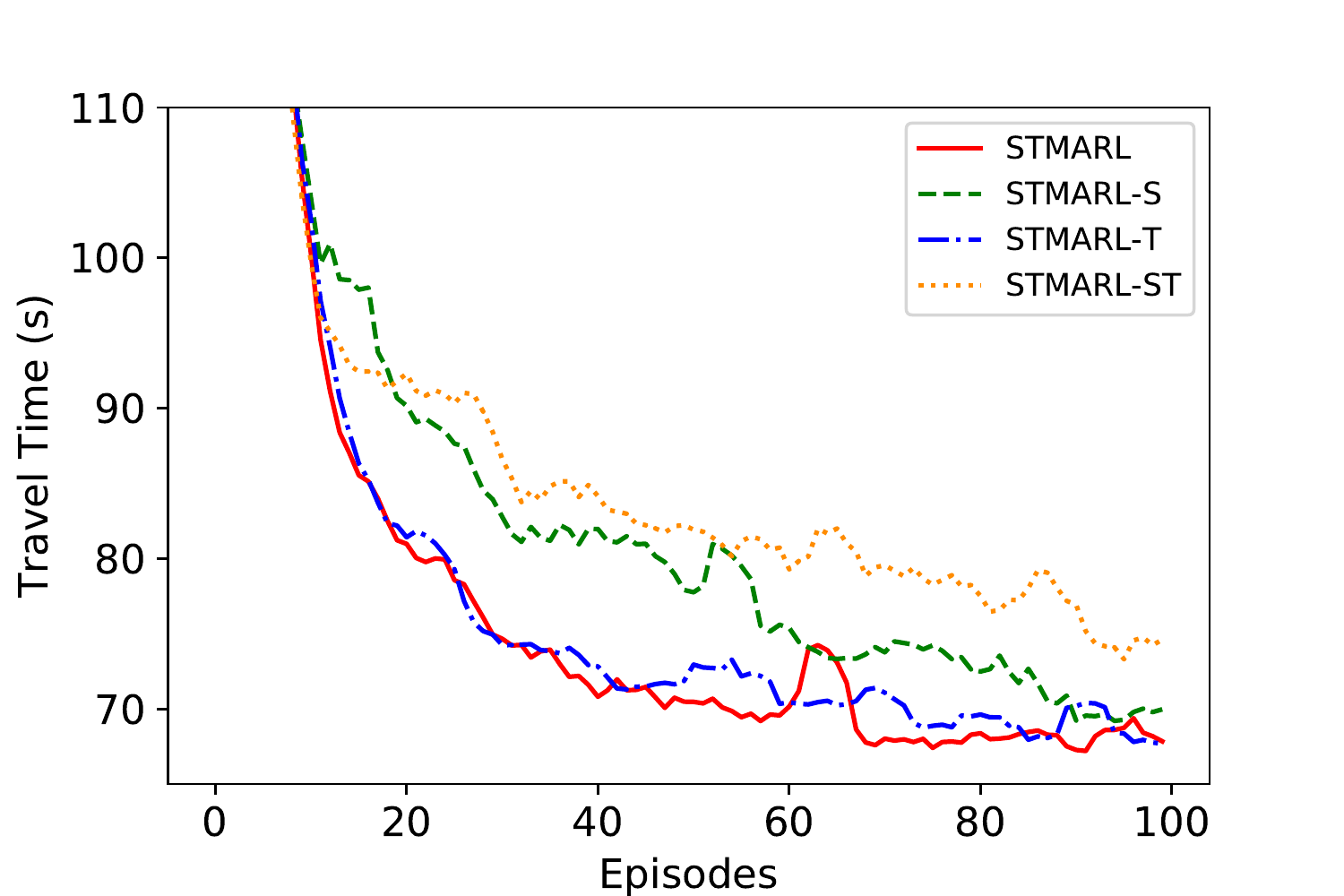}}\label{fig:train_curve_hf}
  \caption{Training curve of our model variants across different datasets. Curves are smoothed with a moving average of five points.}\label{fig:learning curve}
\end{figure*}

Transportation methods are listed as follows:
\begin{itemize}
    \item \textbf{Fixed-time Control: (Fixed-time)~\cite{miller1963settings}}, which uses a pre-defined plan for traffic light control.
    \item \textbf{MaxPressure~\cite{varaiya2013max}}, It is the state-of-the-art transportation method, which greedily choose the phase with the maximum pressure to optimize network-level traffic light control.
\end{itemize}

The following reinforcement learning methods are compared:
\begin{itemize}
    \item \textbf{Max-Plus Coordination for Urban Traffic Control (Max-Plus)~\cite{kuyer2008multiagent}}, which uses max-plus~\cite{kok2006collaborative} algorithm to learn the optimal joint action based on a constructed undirected coordination graph. 
    \item \textbf{Neighbor RL~\cite{arel2010reinforcement}}, which concatenate their neighbor's observation information into their own state representation. It does not distinguish different neighboring states.
    \item \textbf{GCN-lane~\cite{nishi2018traffic}}, which uses graph convolution neural network to extract traffic features of distant roads. Its graph is constructed on lane-level where each lane is regarded as a node and vehicle traffic movement connected two lanes is represented as an edge.
    \item \textbf{GCN-inter}, which constructs its graph on intersection-level where each intersection is regarded as a node and the road connected two intersections is represented as an edge.
    \item \textbf{Colight~\cite{colight}}, which is a recent method using graph attention network for multi-intersection traffic light control. This method determines the agent's neighbors using rules and the number of each agent's neighbors is predefined~\footnote{Colight is tested based on the code https://github.com/wingsweihua/colight}. 
\end{itemize}

  Besides, the following variations of the proposed STMARL model are compared:
\begin{itemize}
    \item \textbf{STMARL-ST}, which is the base independent DQN method with shared parameters among the agents. Specifically, the first encoder layer is separated to handle heterogeneous input information, and parameters for other layers are shared.
    \item \textbf{STMARL-T}, which does not learn temporal dependency and only incorporates the spatial structure information for iterative relational reasoning.
    \item \textbf{STMARL-S}, which only learns the temporal dependency to incorporate the historical traffic information without incorporating spatial structure dependency.
\end{itemize}

\begin{figure*}[htbp]
    \centering
    \subfigure[Performance of STMARL model with different $\Delta t$.]{\includegraphics[width=0.4\textwidth]{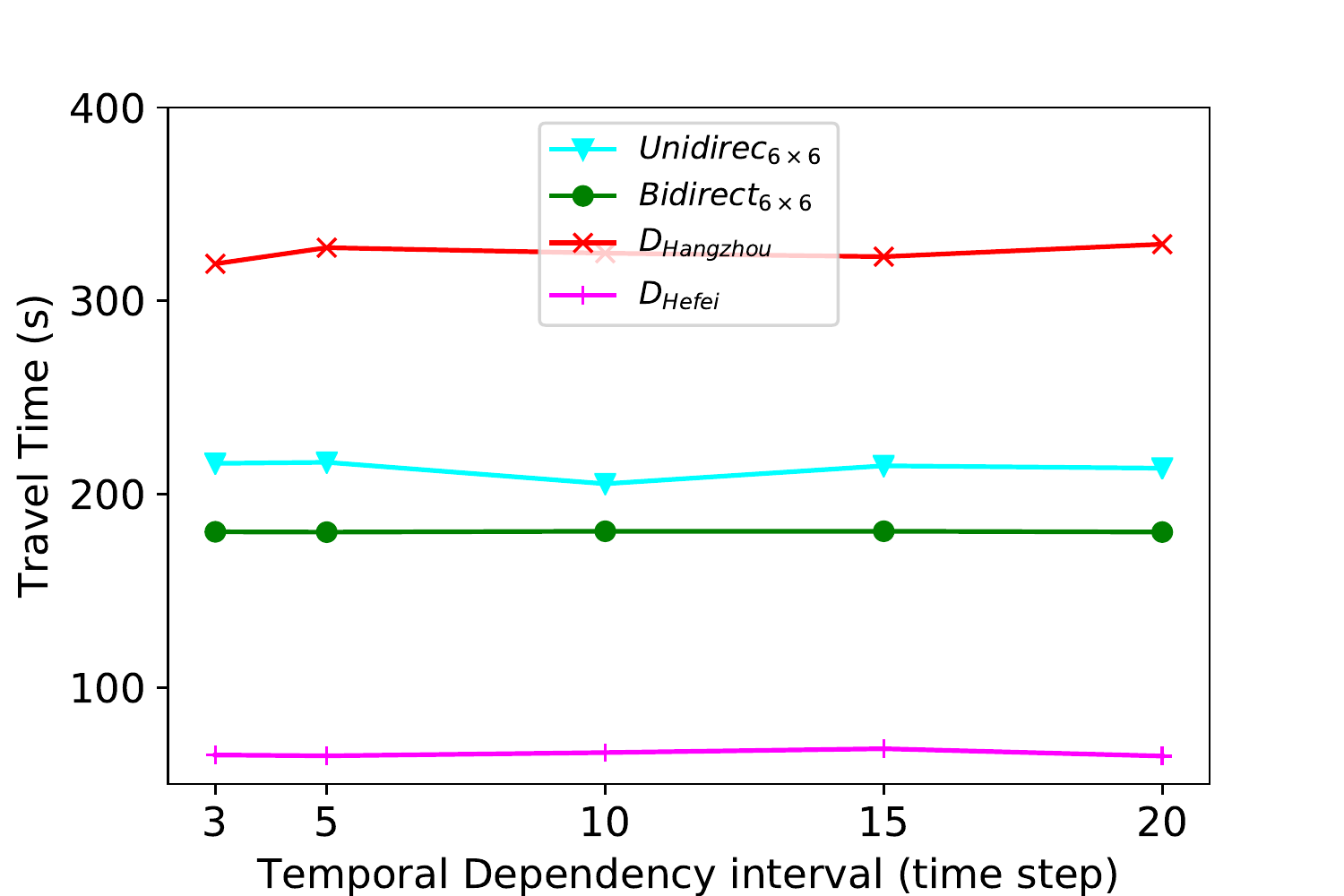}}\label{fig:time_para}
    \subfigure[Running time of STMARL model with different $\Delta t$.]{\includegraphics[width=0.4\textwidth]{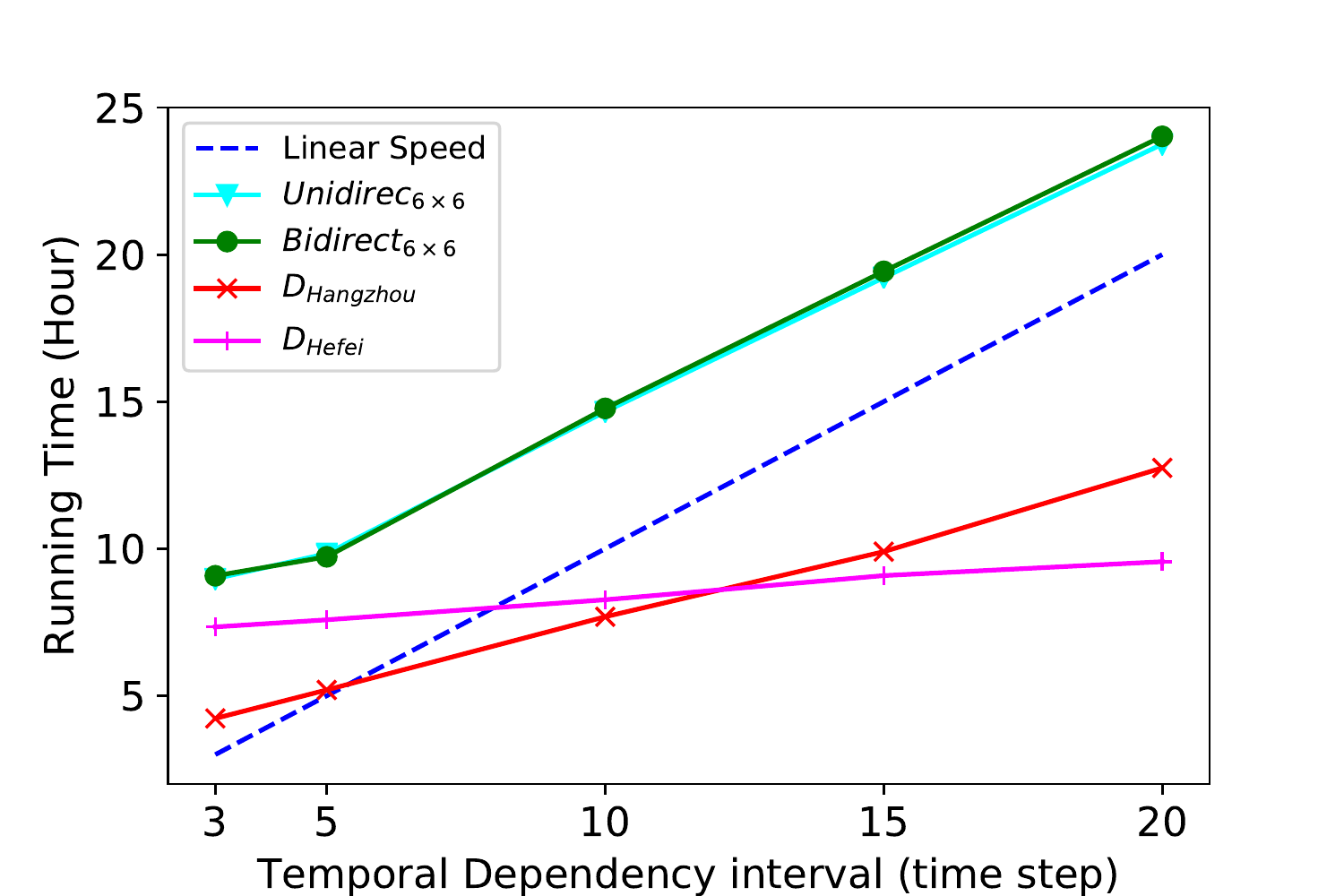}}\label{fig:time_cost}
    \caption{Influence of Temporal Dependency Interval $\Delta t$.}\label{fig:influence_t}
\end{figure*}

For better illustration, we summarize the characteristics of learning-based methods in Table~\ref{table:model_cmpare}.
For the reinforcement learning methods, we trained the model for 100 episodes and then tested with epsilon $\epsilon=0$.
\begin{figure*}[t]
    \fontsize{8}{10}\selectfont
    \centering
    \subfigure[Average arrival rate for intersection 2 on November 6th (Tuesday).]{\includegraphics[width=0.35\textwidth]{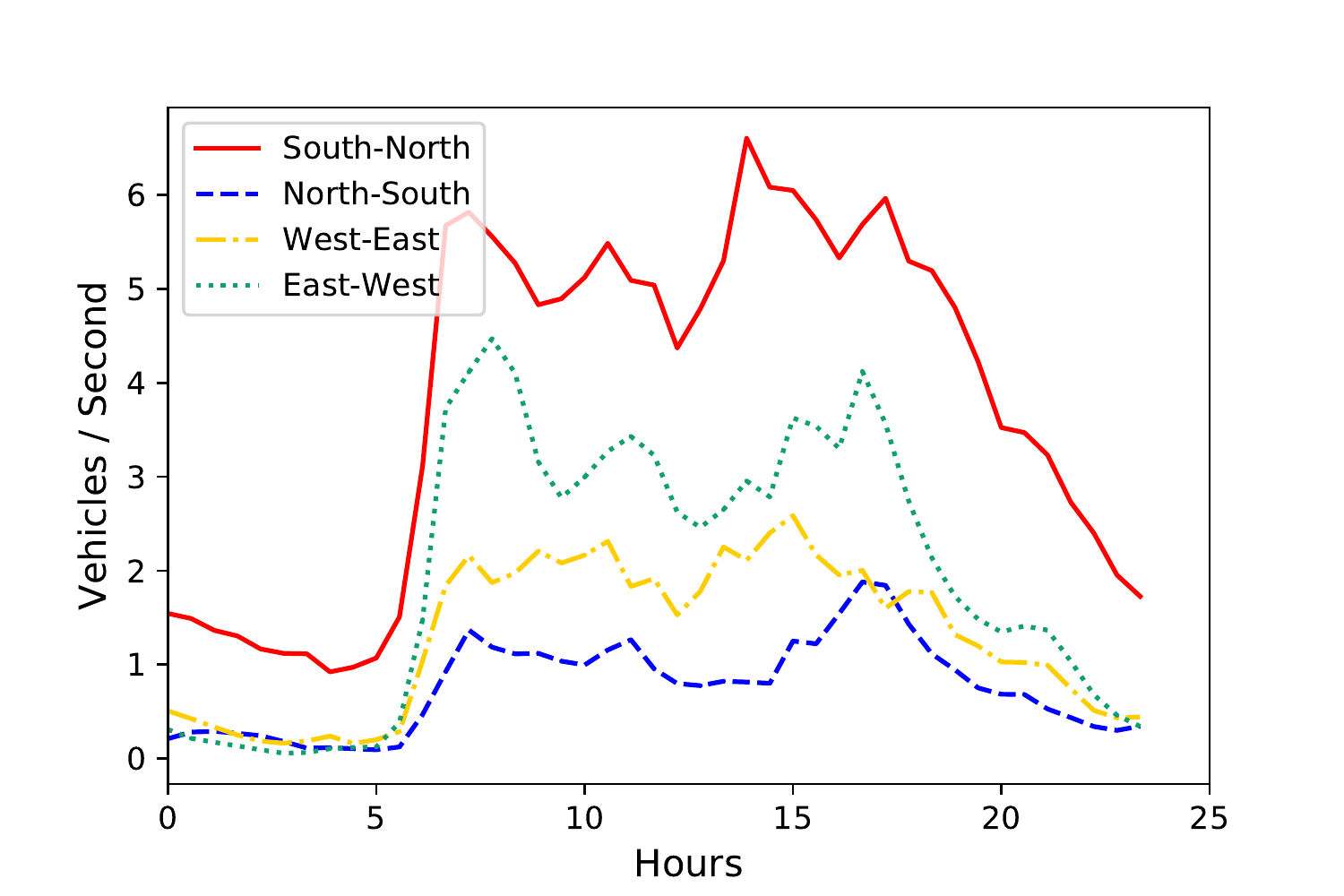}}\label{fig:06_1_arrival_rate}
    \subfigure[Average arrival rate for intersection 2 on November 10th (Saturday).]{\includegraphics[width=0.35\textwidth]{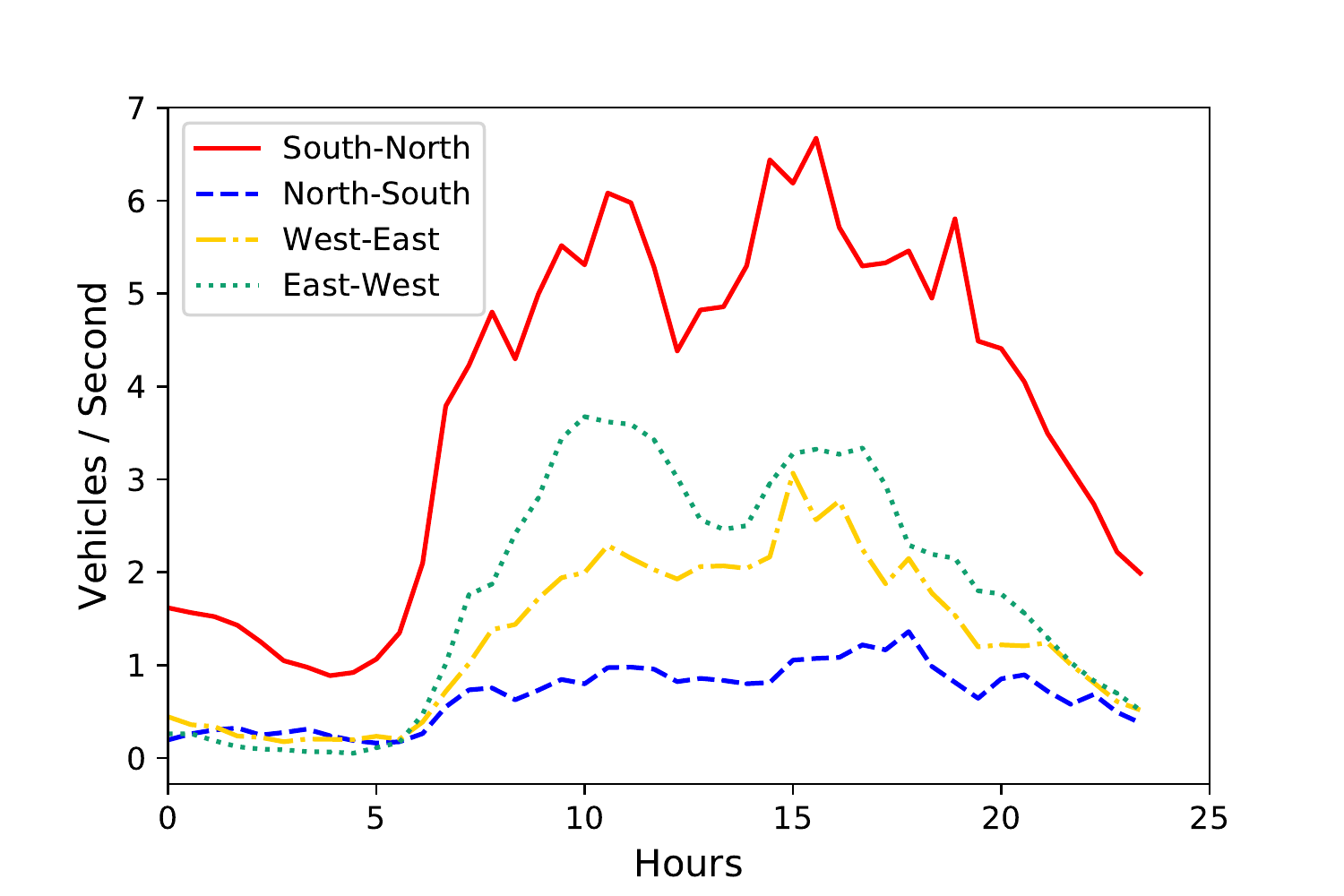}}\label{fig:11_1_arrival_rate}
    \subfigure[Learned attention weights for intersection 2 on November 6th (Tuesday).]{\includegraphics[width=0.35\textwidth]{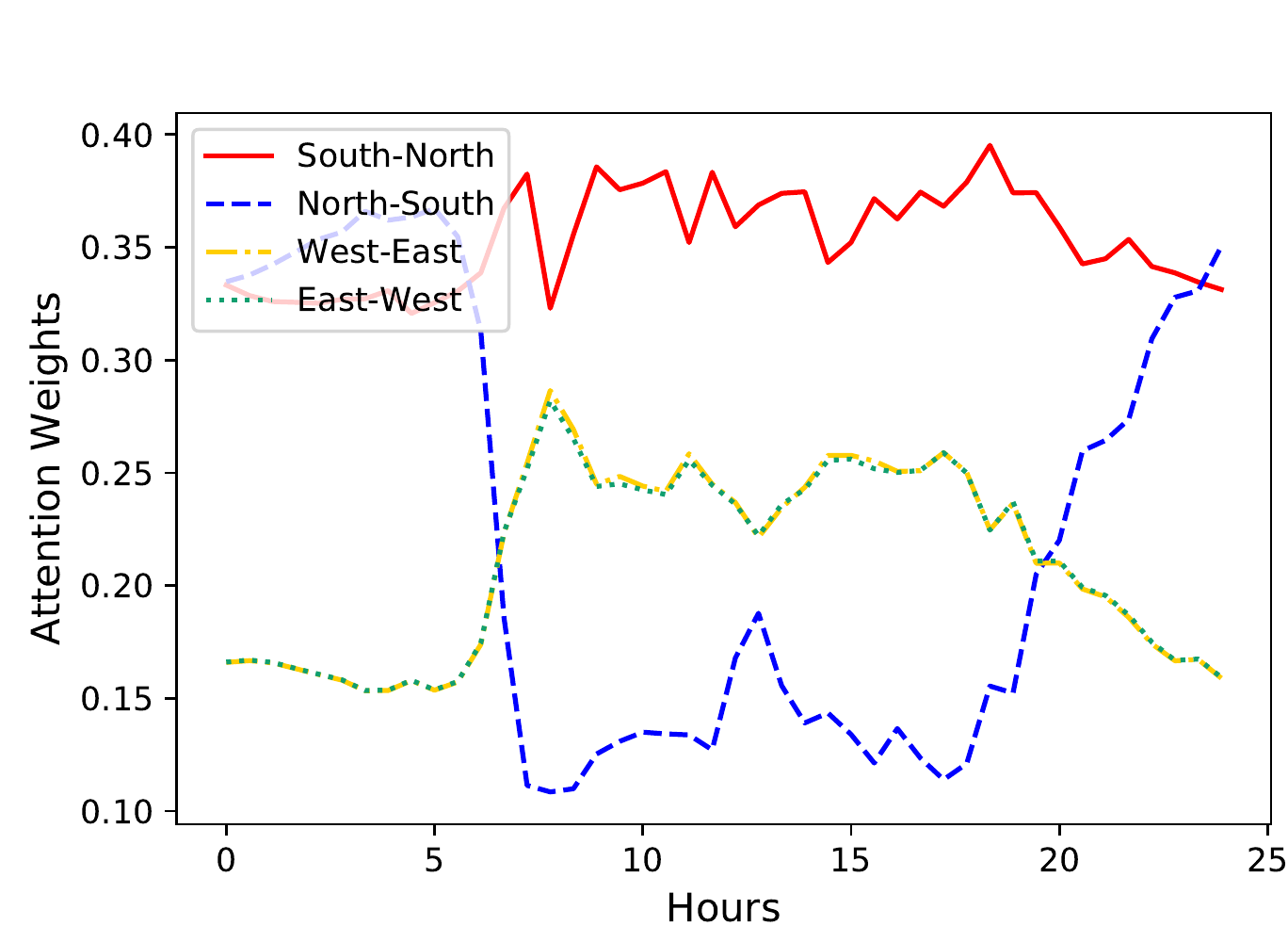}}\label{fig:06_1_att}
    \subfigure[Learned attention weights for intersection 2 on November 10th (Saturday).]{\includegraphics[width=0.35\textwidth]{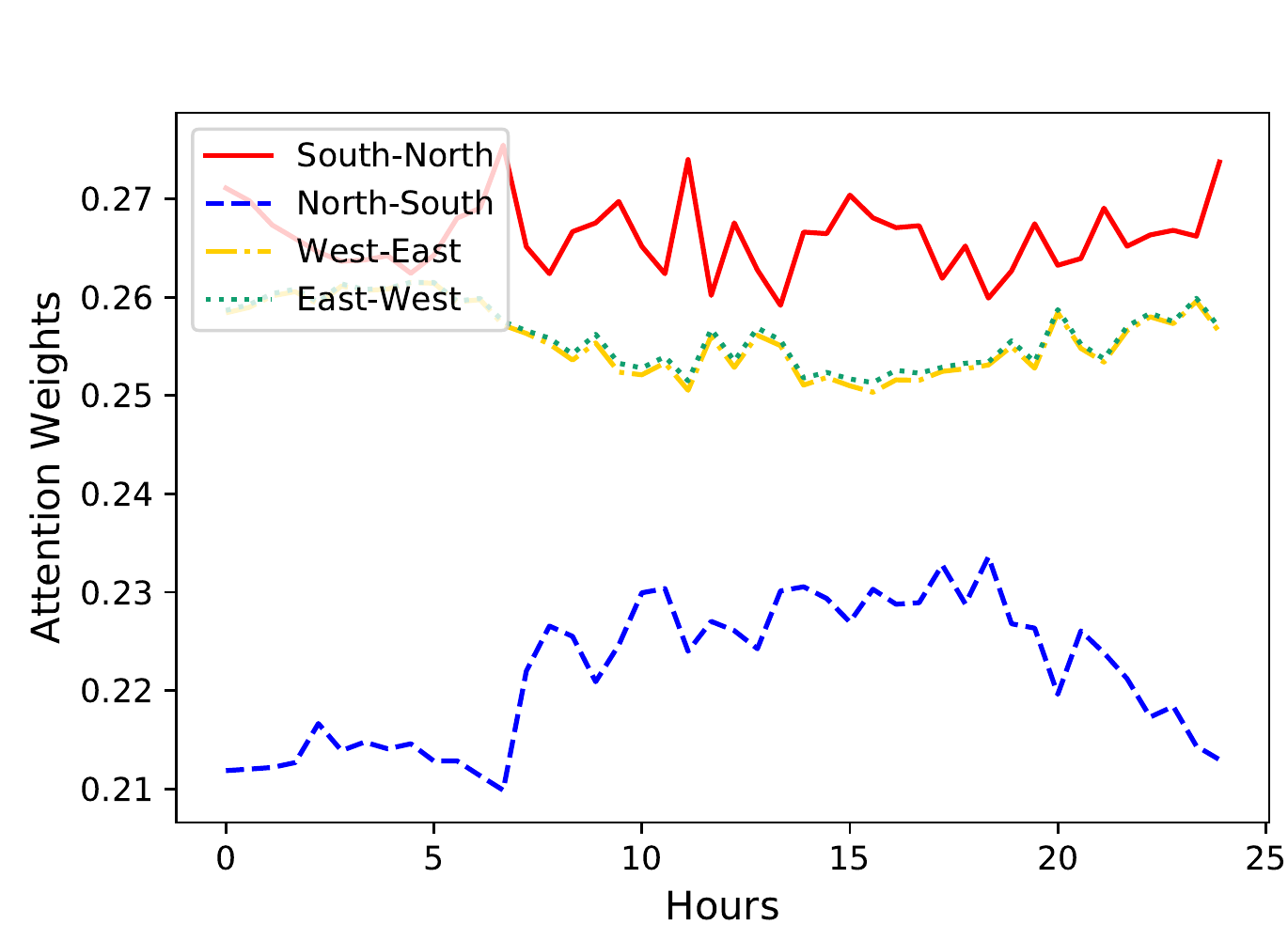}}\label{fig:11_1_att}
    \caption{Average arriving vehicles ((a), (b)) and learned attention weights ((c), (d)) by STMARL model on the four incoming edges (e.g., South-North means the edge direction is from South to North connected to this intersection) of intersection 2 (node TL 2 in Figure~\ref{fig:tsl_graph}) On November 6th (Tuesday) and November 10th (Saturday), 2018, Hefei. 
   }\label{fig:att_weights}
\end{figure*}

\begin{figure*}[t]
    \centering
    \subfigure[Green wave on early morning.]{\includegraphics[width=0.3\textwidth]{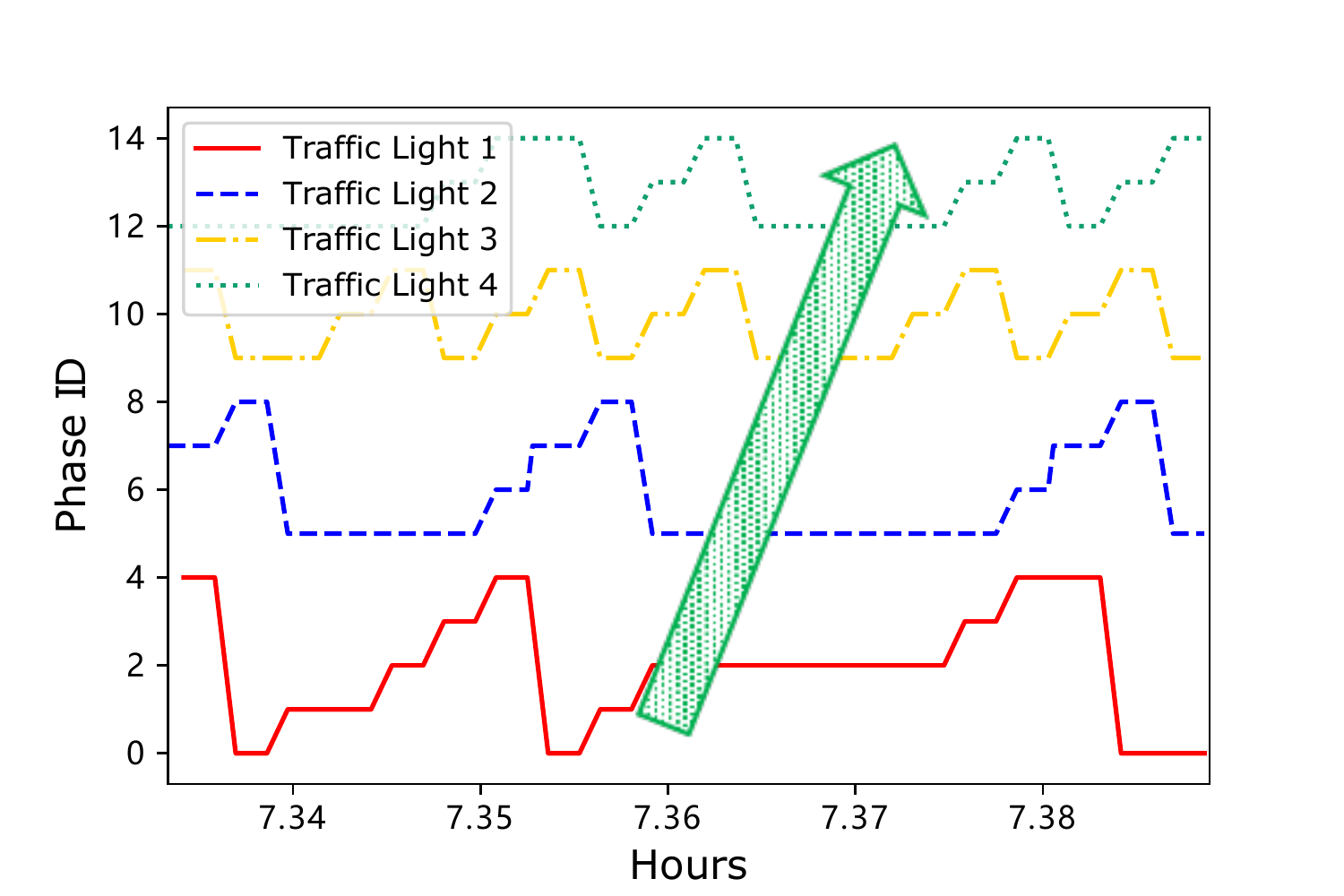}}\label{fig:g_1_1}
    \subfigure[Green wave on middle noon.]{\includegraphics[width=0.3\textwidth]{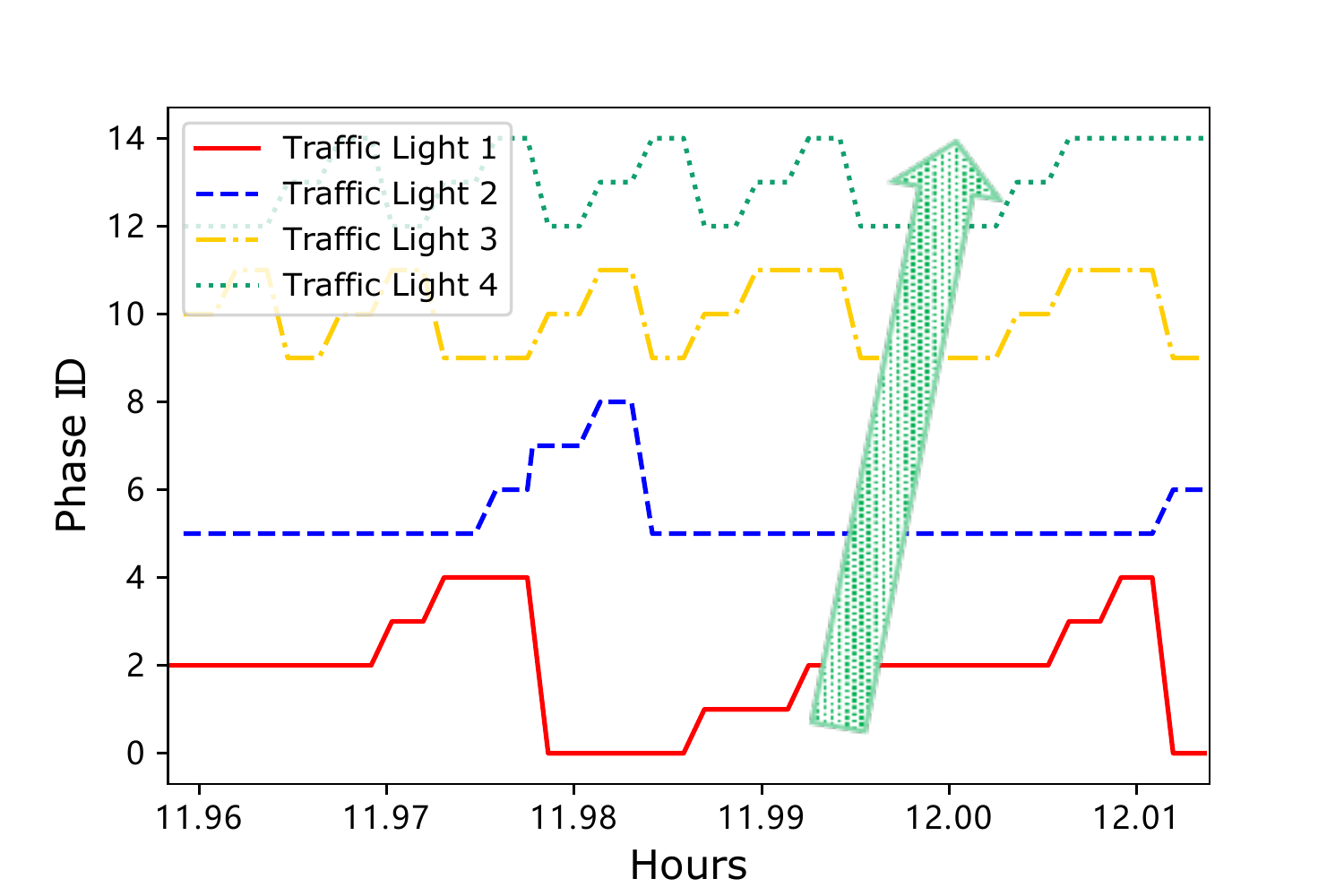}}
    \subfigure[Green wave on latter night.]{\includegraphics[width=0.3\textwidth]{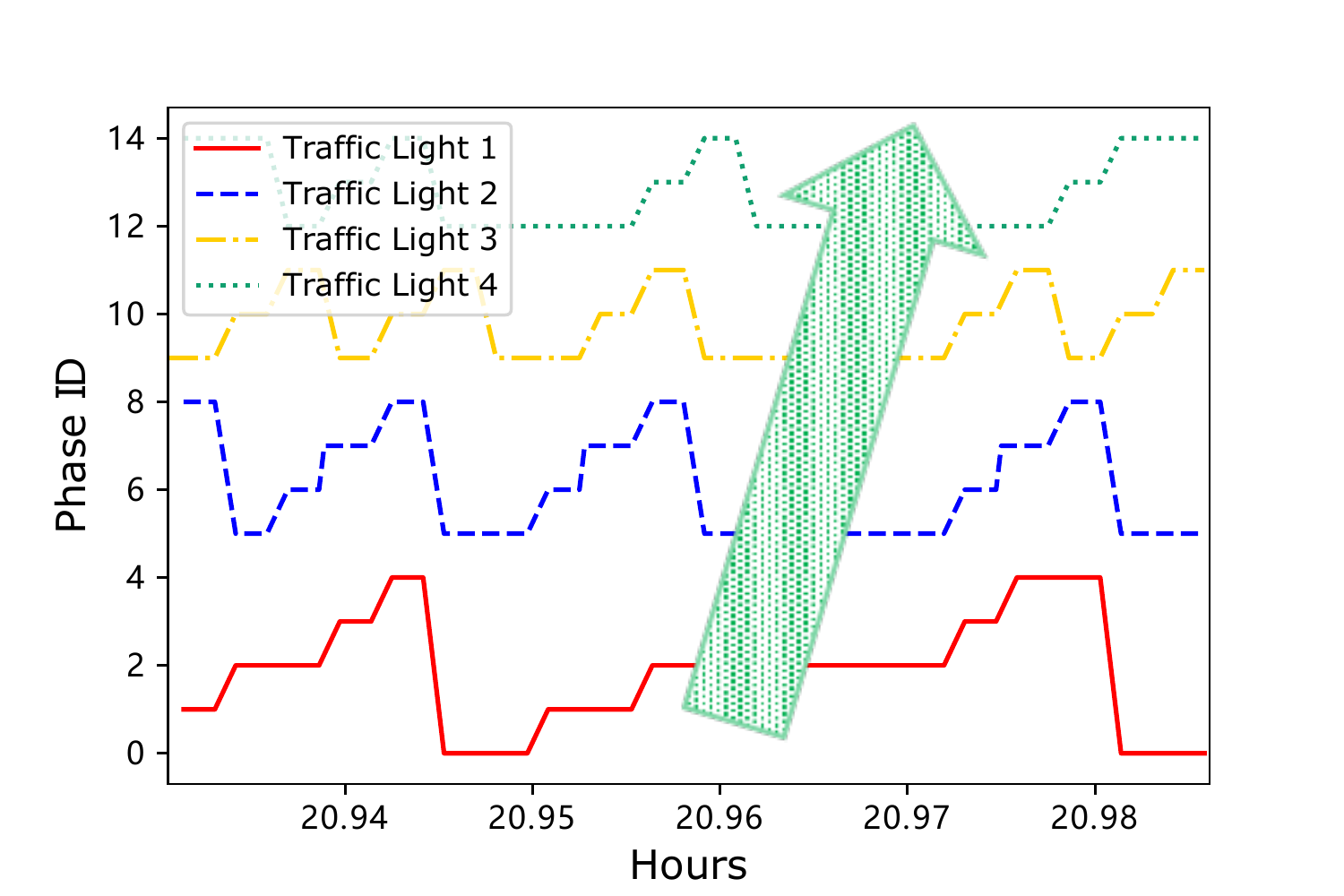}}
    \subfigure[Number of vehicles approaching on early morning.]{\includegraphics[width=0.3\textwidth]{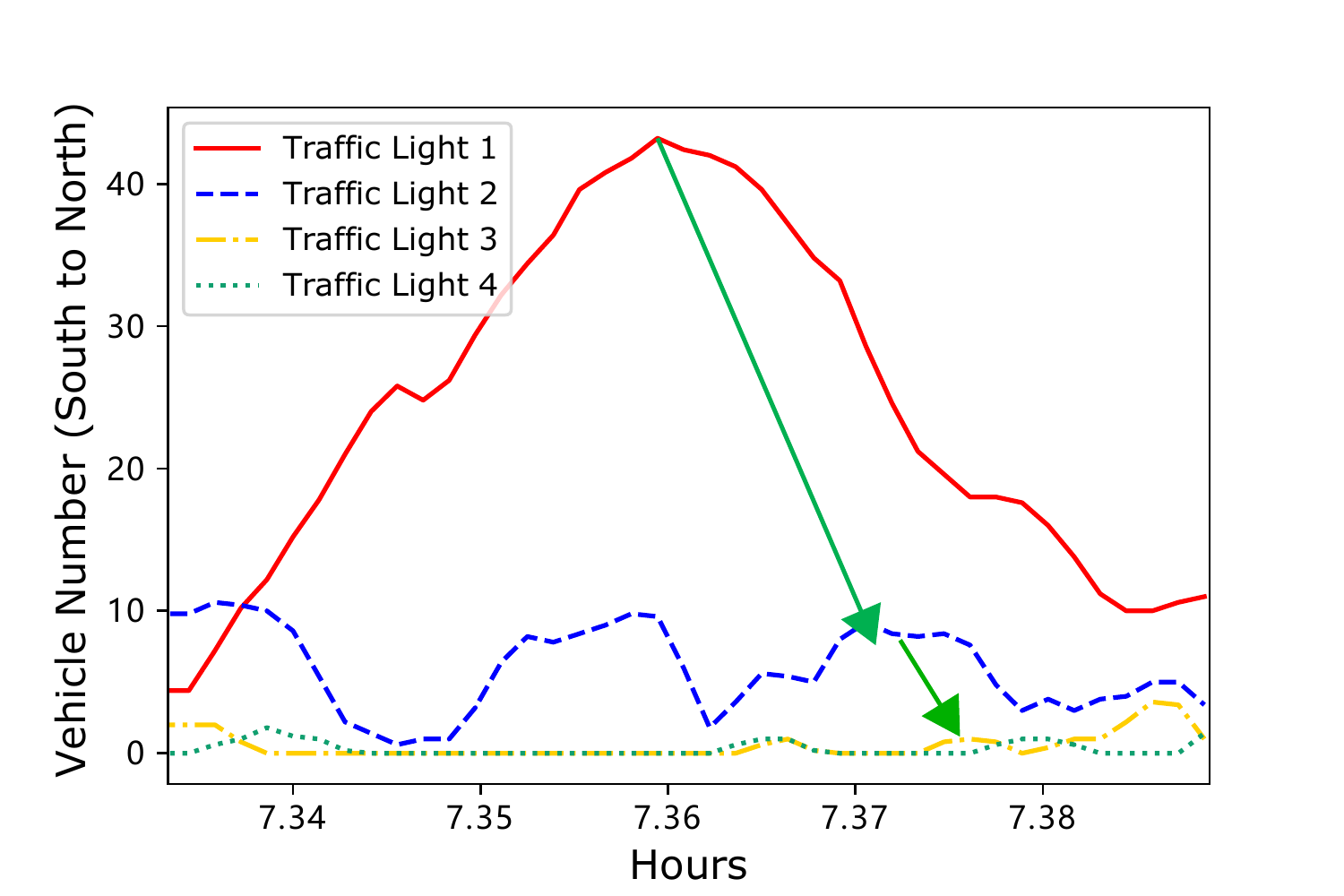}}\label{fig:g_1_2}
    \subfigure[Number of Vehicles approaching on middle noon.]{\includegraphics[width=0.3\textwidth]{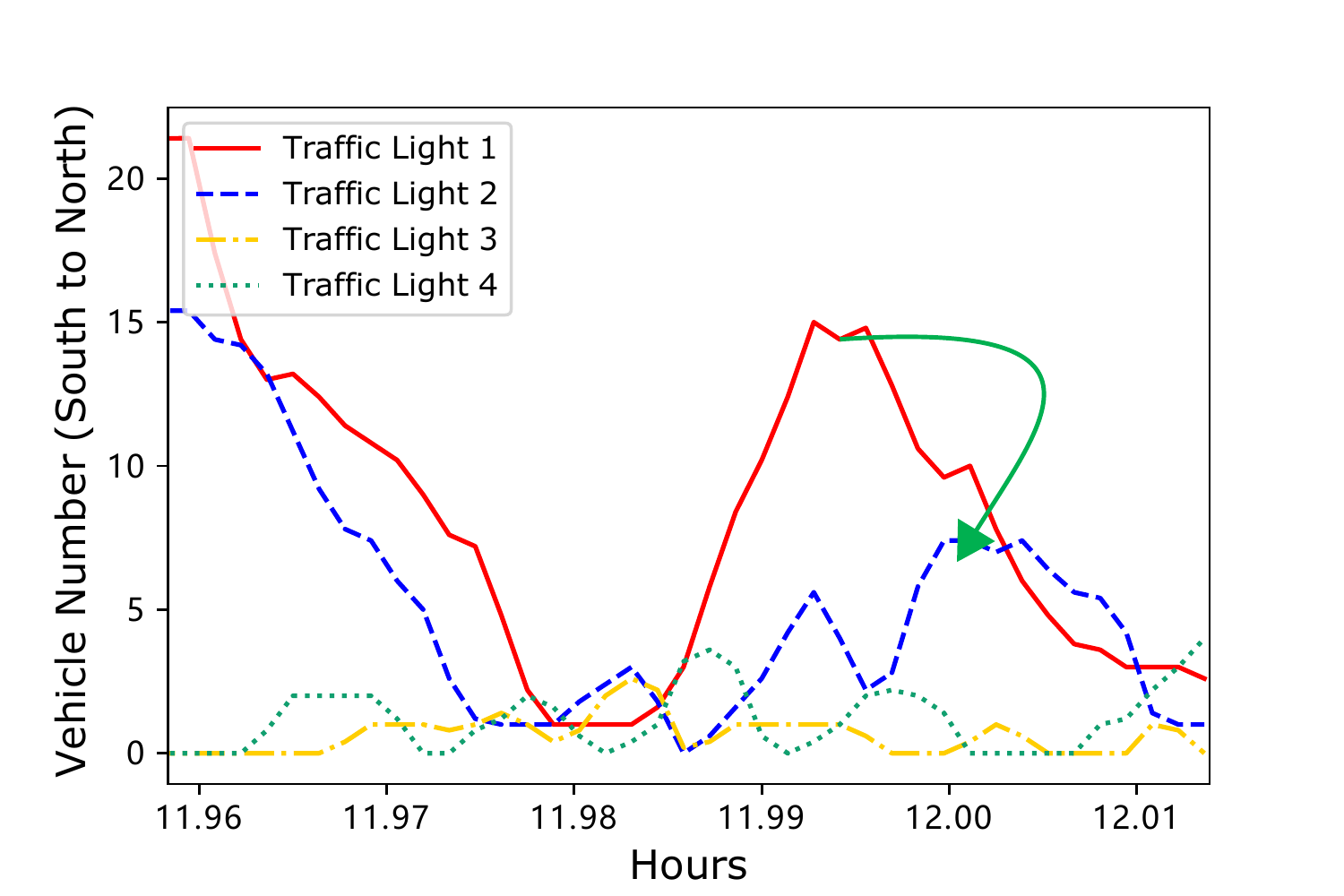}}
    \subfigure[Number of vehicles approaching on latter night.]{\includegraphics[width=0.3\textwidth]{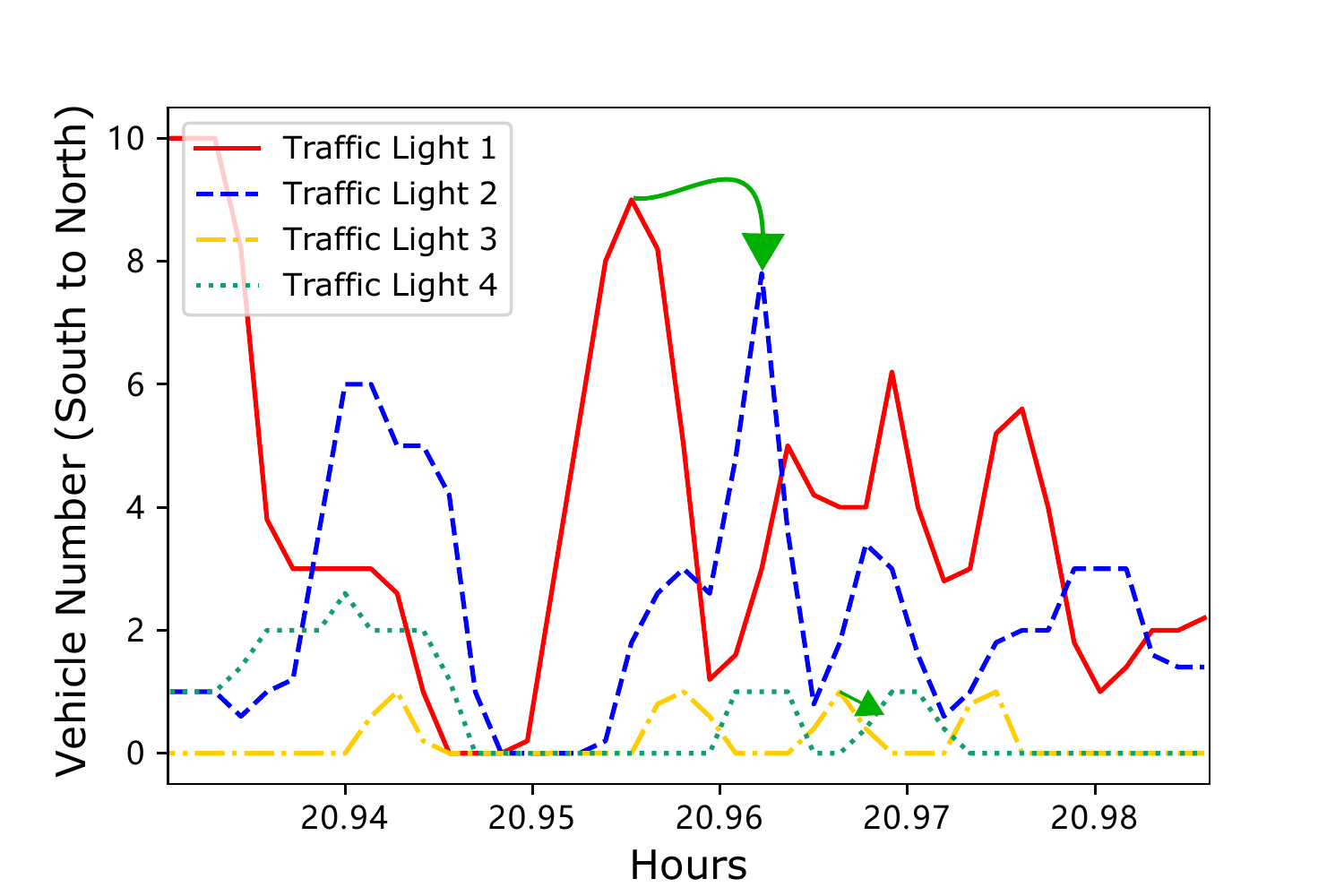}}
    \caption{The emerging green waves show in different time periods on November 6th, 2018, Hefei, in which the waves pointed by the green arrows indicate the green phases of traffic lights coordinated in South to North direction.}\label{fig:green_wave}
\end{figure*}

\subsection{Experimental Results}\label{sec:ex_result}
\subsubsection{Overall Result}
In this section, we compared the proposed method with the baseline methods on both synthetic and real-world datasets. The performance is shown in Table~\ref{table:main_result}. We can observe that our proposed STMARL method significantly outperforms all the baseline methods in all datasets. 

For the comparison of performance across different datasets, we observed that the performance gap becomes significantly larger when the traffic pattern changes from synthetic to real. For example, STMARL outperforms the best baseline by 20.6\% in dataset $D_{Hefei}$.

Compared with transportation methods, we found that the gap between STMARL and transportation methods becomes larger when traffic pattern changes from synthetic to real. This phenomenon demonstrates the effectiveness of the reinforcement learning methods which change traffic phases adaptively to optimize the long term traffic situation.

Besides, our STMARL model significantly outperforms all the reinforcement learning baseline methods. We observed that Neighbor RL performs well under synthetic unidirectional or bidirectional traffic but fails in the large scale real-world traffic flow especially for $D_{Hangzhou}$. The reason may be that Neighbor RL only consider the one-hop neighbor relationship as well as does not consider the weight of their neighbors to respond to the real dynamic traffic flow. 
It can be also seen that STMARL outperforms Colight by a large margin over all the datasets. This observation indicates that it is effective to model the cooperation structure by leveraging the constructed directional traffic light adjacency graph as well as the temporal dependency. On the contrary, Colight determines each agent's neighbors by rules, and the number of neighbors is fixed, which also ignores modeling traffic flow direction on the graph. What's more,
compared with these two GCN based methods, STMARL is superior in performance. As STMARL constructed the graph on the intersection level, it is more effective to model the relationship among traffic lights than the lane-level graph used by GCN-lane. The reason that GCN-lane fails to learn well in large scale road networks may also be due to its significantly increased graph complexity with a large road network such as the synthetic $6\times6$ road network.
When building its graph on the intersection level, GCN-inter performs better than GCN-lane but fails to learn the dynamic real-world traffic flow as it treats the neighbors equally. These comparisons clearly demonstrate the effectiveness of STMARL by collectively learning spatial-temporal dependency based on the directional traffic light adjacency graph. 

\subsubsection{Ablation Study}
The ablation study of the model component is shown in Table~\ref{table:ablation_study}. We can find that STMARL constantly outperforms all the model variants over the datasets. Particularly, incorporating spatial structure dependency boosts performance more than the temporal dependency in most of the datasets. This demonstrates that learning spatial structure dependency is more important for cooperation among traffic lights to improve performance.
Figure~\ref{fig:learning curve} illustrates the training curves of these model variants across different datasets. It can be observed that incorporating spatial structure dependency greatly improved the convergence speed. Adding temporal dependency will further accelerate convergence and improve performance. These quantitative results clearly demonstrate the effectiveness of jointly learning spatial structure information and temporal dependency for multi-intersection traffic light control.



\subsubsection{Influence of Temporal Dependency Interval $\Delta t$} In this section, we show the sensitiveness of temporal dependency interval $\Delta t$ and the scalability of STMARL with different $\Delta t$.

\noindent\textbf{Sensitiveness.} Figure~\ref{fig:influence_t} (a) shows the performance of STMARL model with different temporal dependency interval $\Delta t$. We observe that STMARL achieves the best performance when $\Delta t = 10, 5, 3, 20$ for $Unidirec_{6 \times 6}$, $Bidirect_{6 \times 6}$, $D_{Hangzhou}$ and $D_{Hefei}$ respectively. These results indicate that a relatively medium time interval $\Delta t$ should be more appropriate to learn the temporal dependency.

\noindent\textbf{Scalability.}
Figure~\ref{fig:influence_t} (b) shows the running time of STMARL model under different temporal dependency interval across different datasets for 100 episodes. It can be observed that the training time of the STMARL scale almost linearly with the increasing $\Delta t$ under different scale road networks. On the real-world datasets $D_{Hangzhou}$ and $D_{Hefei}$, the training of STMARL is efficient across different $\Delta t$, which demonstrates the scalability of STMARL on large scale real-world traffic light control.

\textcolor{black}{\subsubsection{Influence of Hidden Layer Size $h$}}
\begin{figure}[t]
    \centering
    \includegraphics[width=0.45\textwidth]{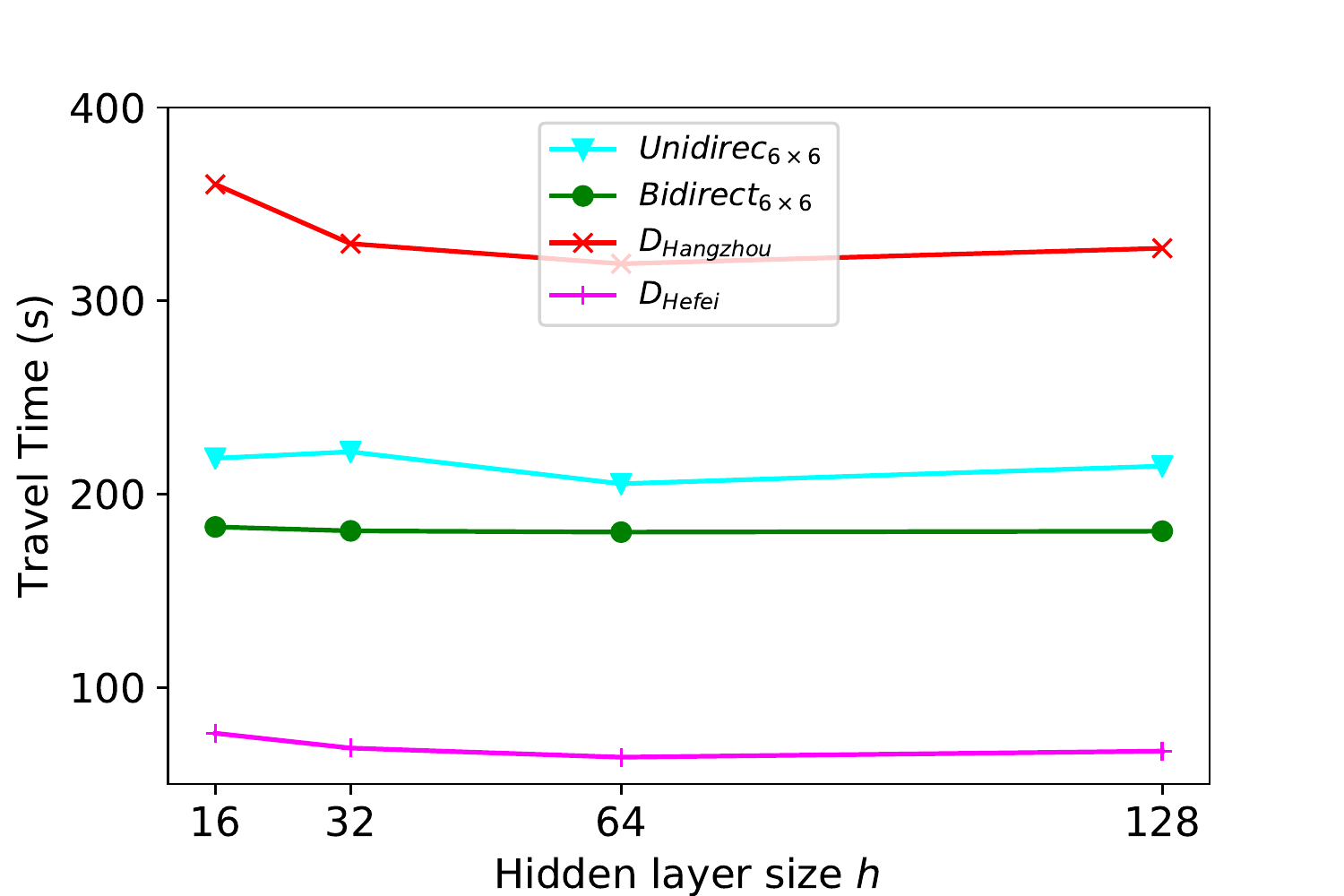}
    \caption{\textcolor{black}{Influence of hidden layer size $h$}.}
    \label{fig:para_hid}
\end{figure}
\textcolor{black}{Figure ~\ref{fig:para_hid} shows the performance of STMARL with different hidden layer seize $h$ across all datasets. We can observe that STMARL achieves the best performance when the hidden layer size is 64 for all the datasets, which indicates a medium time complexity according to Section ~\ref{Sec:time_complexity}.}

\textcolor{black}{\subsubsection{Fairness of STMARL Method}}
\textcolor{black}{In this section, we discuss the fairness of STMARL method among vehicles. As shown in Table ~\ref{tab:result_std}, the standard deviation of the travel time among vehicles for STMARL method is the smallest compared with the baseline methods in all the datasets. This result demonstrates that our STMARL method is fair among vehicles.}
\begin{table}[t]
\fontsize{8}{10}\selectfont
\centering
\caption{\textcolor{black}{Standard deviation of the travel time among vehicles in an area.}}\label{tab:result_std}
\scalebox{0.95}{\begin{tabular}{c|cccc}
\hline
Methods  & $Unidirec_{6 \times 6}$                   & $Bidirect_{6 \times 6}$             & $D_{Hangzhou}$       & $D_{Hefei}$                   \\ \hline
 Fixed-time &167.28   & 151.09   &520.96   &276.32   \\
 MaxPressure &73.97  &50.37  &447.54 &224.11  \\\hline
 Max-Plus & 804.43  & 975.38  & 1179.10  & 211.17  \\ 
 Neighbor RL &39.65  & 26.97  &226.66  &191.76  \\
 GCN-lane&514.76   & 36.05  & 740.16  &231.20  \\
 GCN-inter&62.74  & 35.88  &316.33  &214.61  \\
 Colight&273.92  & 194.84  &394.10  &655.38 \\\hline
 \textbf{STMARL}&$\textbf{36.10}$  &$\textbf{26.14}$ &$\textbf{208.64}$  &$\textbf{174.47}$  \\ \hline
\end{tabular}}
\end{table}
\subsubsection{Qualitative Study} \label{sec:qualitative}
In this section, we provide further analysis of the learned attention weights to the adjacent nodes and the emergence of the green wave learned by the STMARL model in the real-world dataset $D_{Hefei}$ for full-day analysis.

\noindent\textbf{Interpretation of Attention Weights.}
In this section, we analyze the learned attention weights for the traffic light agent and take the Traffic Light agent 2 as an example. Figure~\ref{fig:att_weights} (c)(d) shows the learned attention weights in four incoming edges from its neighbors on Tuesday and Saturday. We also show the average number of approaching vehicles in corresponding edges in Figure~\ref{fig:att_weights} (a)(b). It can be observed that the learned attention weights keep pace with the dynamic number of vehicles in the corresponding edges. For example, in different peak hours, the attention weights in the corresponding edge directions also become larger.

Moreover, in Figure~\ref{fig:att_weights} (c), the attention weight of the South-North direction is the largest compared to the other three directions in most cases, which corresponds to the largest arriving vehicle numbers as shown in Figure~\ref{fig:att_weights} (a). Similar rule could be found between Figure~\ref{fig:att_weights} (d) and Figure~\ref{fig:att_weights} (b). Therefore, the larger attention weights make the Traffic Light agent~2 care more about the downstream traffic situation which may spill into intersection~2. As a result, the Traffic Light agent~2 was influenced by the decision of Traffic Light agent~1. Along this line, the coordination among Traffic Light agent~2 and agent~1 is crucial so as not to cause severe traffic congestion to Intersection~2 due to the large traffic flow from intersection~1. Therefore, the larger attention weights in the edge direction may indicate that it's more necessary to coordinate among these two~agents.

\noindent\textbf{Cooperation for Green Wave.}
The green wave occurs when a series of traffic lights are coordinated to allow continuous traffic flow over several intersections along one main direction. Therefore, it can be used to test the coordination mechanism learned by multiple traffic lights. Figure~\ref{fig:green_wave} shows the dynamics of traffic light phase learned by our model (Figure~\ref{fig:green_wave} (a)(b)(c)) and the corresponding number of approaching cars along South to North direction (Figure~\ref{fig:green_wave} (d)(e)(f)) which shows the emergence of \emph{green wave} phenomenon. It can be observed that from Figure~\ref{fig:green_wave} (a)(b)(c), in these three time periods, there exists a green wave, i.e., green arrow, where all the four traffic light agents coordinated their traffic phases (current green phase in South-North direction) to allow fast traveling for the approaching cars.  Figure~\ref{fig:green_wave} (d)(e)(f) shows that the green wave significantly accelerates the traffic flow by reducing the maximum number of vehicles approaching one intersection (e.g., the intersection controlled by Traffic Light 1). It also shows the peak of vehicle number moves from intersection 1 to intersection 2 along the green wave direction, which indicates the fast-moving of traffic flow. Therefore, the green wave demonstrates that the STMARL model can learn coordination policy to reduce traffic congestion at an integral level.


\section{Conclusions}
\label{sec:conclusion}
In this paper, we proposed the Spatio-Temporal Multi-Agent Reinforcement Learning (STMARL) model for multi-intersection traffic light control. The proposed STMARL method can leverage the spatial structure in the real world to facilitate coordination among multiple traffic lights. Moreover, it also considers the historical traffic information for current decision making.
Specifically, we first construct the traffic light adjacency graph based on the spatial structure among traffic lights. Then, historical traffic records will be integrated with current traffic status via Recurrent Neural Network structure. Moreover, based on the temporally-dependent traffic information, we design a Graph Neural Network based model to represent relationships among multiple traffic lights, and the decision for each traffic light will be made in a distributed way by deep Q-learning method. Experiments on both synthetic and real-world datasets have demonstrated the effectiveness of our STMARL framework, which also provides an insightful understanding of the influence mechanism among multi-intersection traffic lights.

\section*{Acknowledgement}
This research was partially supported by grants from the National Key Research and Development Program of China (Grant No.2018YFB1402600), and the National Natural Science Foundation of China (Grant No.91746301, 71531001, 61703386, U1605251).




\bibliographystyle{IEEEtran}
\bibliography{references}

\begin{thebibliography}{10}
\providecommand{\url}[1]{#1}
\csname url@samestyle\endcsname
\providecommand{\newblock}{\relax}
\providecommand{\bibinfo}[2]{#2}
\providecommand{\BIBentrySTDinterwordspacing}{\spaceskip=0pt\relax}
\providecommand{\BIBentryALTinterwordstretchfactor}{4}
\providecommand{\BIBentryALTinterwordspacing}{\spaceskip=\fontdimen2\font plus
\BIBentryALTinterwordstretchfactor\fontdimen3\font minus
  \fontdimen4\font\relax}
\providecommand{\BIBforeignlanguage}[2]{{%
\expandafter\ifx\csname l@#1\endcsname\relax
\typeout{** WARNING: IEEEtran.bst: No hyphenation pattern has been}%
\typeout{** loaded for the language `#1'. Using the pattern for}%
\typeout{** the default language instead.}%
\else
\language=\csname l@#1\endcsname
\fi
#2}}
\providecommand{\BIBdecl}{\relax}
\BIBdecl

\bibitem{traficjamcost}
\BIBentryALTinterwordspacing
CityLab, ``Traffic's mind-boggling economic toll,'' 2018. [Online]. Available:
  \url{https://www.citylab.com/transportation/2018/02/traffics-mind-boggling-economic-toll/552488/}
\BIBentrySTDinterwordspacing

\bibitem{xu2019exploring}
T.~Xu, H.~Zhu, H.~Xiong, H.~Zhong, and E.~Chen, ``Exploring the social learning
  of taxi drivers in latent vehicle-to-vehicle networks,'' \emph{IEEE
  Transactions on Mobile Computing}, 2019.

\bibitem{sommer2010bidirectionally}
C.~Sommer, R.~German, and F.~Dressler, ``Bidirectionally coupled network and
  road traffic simulation for improved {IVC} analysis,'' \emph{IEEE
  Transactions on mobile computing}, vol.~10, pp. 3--15, 2010.

\bibitem{wu2019block}
C.~Wu, A.~Pozdnukhov, and A.~M. Bayen, ``Block simplex signal recovery:
  Methods, trade-offs, and an application to routing,'' \emph{IEEE Transactions
  on Intelligent Transportation Systems}, vol.~21, pp. 1547--1559, 2019.

\bibitem{cabannes2019regrets}
T.~Cabannes, M.~Sangiovanni, A.~Keimer, and A.~M. Bayen, ``Regrets in routing
  networks: Measuring the impact of routing apps in traffic,'' \emph{ACM
  Transactions on Spatial Algorithms and Systems (TSAS)}, vol.~5, pp. 1--19,
  2019.

\bibitem{li2020competitive}
S.~Li, J.~Zhou, T.~Xu, H.~Liu, X.~Lu, and H.~Xiong, ``Competitive analysis for
  points of interest,'' in \emph{Proceedings of the 26th ACM SIGKDD
  International Conference on Knowledge Discovery \& Data Mining}, 2020, pp.
  1265--1274.

\bibitem{zhang2020semi}
W.~Zhang, H.~Liu, Y.~Liu, J.~Zhou, and H.~Xiong, ``Semi-supervised hierarchical
  recurrent graph neural network for city-wide parking availability
  prediction,'' in \emph{Proceedings of the AAAI Conference on Artificial
  Intelligence}, vol.~34, no.~01, 2020, pp. 1186--1193.

\bibitem{miller1963settings}
A.~J. Miller, ``Settings for fixed-cycle traffic signals,'' \emph{Journal of
  the Operational Research Society}, vol.~14, pp. 373--386, 1963.

\bibitem{yin2016traffic}
B.~Yin, M.~Dridi, and A.~El~Moudni, ``Traffic network micro-simulation model
  and control algorithm based on approximate dynamic programming,'' \emph{IET
  Intelligent Transport Systems}, vol.~10, pp. 186--196, 2016.

\bibitem{cools2013self}
S.-B. Cools, C.~Gershenson, and B.~D\'Hooghe, ``Self-organizing traffic lights:
  A realistic simulation,'' in \emph{Advances in applied self-organizing
  systems}.\hskip 1em plus 0.5em minus 0.4em\relax Springer, 2013, pp. 45--55.

\bibitem{wei2018intellilight}
H.~Wei, G.~Zheng, H.~Yao, and Z.~Li, ``Intellilight: A reinforcement learning
  approach for intelligent traffic light control,'' in \emph{Proceedings of the
  24th ACM SIGKDD International Conference on Knowledge Discovery \& Data
  Mining}.\hskip 1em plus 0.5em minus 0.4em\relax ACM, 2018, pp. 2496--2505.

\bibitem{mannion2016experimental}
P.~Mannion, J.~Duggan, and E.~Howley, ``An experimental review of reinforcement
  learning algorithms for adaptive traffic signal control,'' in \emph{Autonomic
  Road Transport Support Systems}.\hskip 1em plus 0.5em minus 0.4em\relax
  Springer, 2016, pp. 47--66.

\bibitem{watter2015embed}
M.~Watter, J.~Springenberg, J.~Boedecker, and M.~Riedmiller, ``Embed to
  control: A locally linear latent dynamics model for control from raw
  images,'' in \emph{Advances in neural information processing systems}, 2015,
  pp. 2746--2754.

\bibitem{busoniu2008comprehensive}
L.~Busoniu, R.~Babuska, and B.~De~Schutter, ``A comprehensive survey of
  multiagent reinforcement learning,'' \emph{IEEE Transactions on Systems, Man,
  And Cybernetics-Part C: Applications and Reviews, 38 (2), 2008}, 2008.

\bibitem{kuyer2008multiagent}
L.~Kuyer, S.~Whiteson, B.~Bakker, and N.~Vlassis, ``Multiagent reinforcement
  learning for urban traffic control using coordination graphs,'' in
  \emph{Joint European Conference on Machine Learning and Knowledge Discovery
  in Databases}.\hskip 1em plus 0.5em minus 0.4em\relax Springer, 2008, pp.
  656--671.

\bibitem{velickovic2017graph}
\BIBentryALTinterwordspacing
P.~Velickovic, G.~Cucurull, A.~Casanova, A.~Romero, P.~Li{\`{o}}, and
  Y.~Bengio, ``Graph attention networks,'' in \emph{6th International
  Conference on Learning Representations, {ICLR} 2018, Vancouver, BC, Canada,
  April 30 - May 3, 2018, Conference Track Proceedings}.\hskip 1em plus 0.5em
  minus 0.4em\relax OpenReview.net, 2018. [Online]. Available:
  \url{https://openreview.net/forum?id=rJXMpikCZ}
\BIBentrySTDinterwordspacing

\bibitem{el2013multiagent}
S.~El-Tantawy, B.~Abdulhai, and H.~Abdelgawad, ``Multiagent reinforcement
  learning for integrated network of adaptive traffic signal controllers
  {(MARLIN-ATSC)}: methodology and large-scale application on downtown
  toronto,'' \emph{IEEE Transactions on Intelligent Transportation Systems},
  vol.~14, pp. 1140--1150, 2013.

\bibitem{khamis2014adaptive}
M.~A. Khamis and W.~Gomaa, ``Adaptive multi-objective reinforcement learning
  with hybrid exploration for traffic signal control based on cooperative
  multi-agent framework,'' \emph{Engineering Applications of Artificial
  Intelligence}, vol.~29, pp. 134--151, 2014.

\bibitem{abdulhai2003reinforcement}
B.~Abdulhai, R.~Pringle, and G.~J. Karakoulas, ``Reinforcement learning for
  true adaptive traffic signal control,'' \emph{Journal of Transportation
  Engineering}, vol. 129, pp. 278--285, 2003.

\bibitem{el2010agent}
S.~El-Tantawy and B.~Abdulhai, ``An agent-based learning towards decentralized
  and coordinated traffic signal control,'' in \emph{13th International IEEE
  Conference on Intelligent Transportation Systems}.\hskip 1em plus 0.5em minus
  0.4em\relax IEEE, 2010, pp. 665--670.

\bibitem{steingrover2005reinforcement}
M.~Steingrover, R.~Schouten, S.~Peelen, E.~Nijhuis, B.~Bakker \emph{et~al.},
  ``Reinforcement learning of traffic light controllers adapting to traffic
  congestion.'' in \emph{BNAIC}.\hskip 1em plus 0.5em minus 0.4em\relax
  Citeseer, 2005, pp. 216--223.

\bibitem{arel2010reinforcement}
I.~Arel, C.~Liu, T.~Urbanik, and A.~Kohls, ``Reinforcement learning-based
  multi-agent system for network traffic signal control,'' \emph{IET
  Intelligent Transport Systems}, vol.~4, pp. 128--135, 2010.

\bibitem{rizzo2019time}
S.~G. Rizzo, G.~Vantini, and S.~Chawla, ``Time critic policy gradient methods
  for traffic signal control in complex and congested scenarios,'' in
  \emph{Proceedings of the 25th ACM SIGKDD International Conference on
  Knowledge Discovery \& Data Mining}.\hskip 1em plus 0.5em minus 0.4em\relax
  ACM, 2019, pp. 1654--1664.

\bibitem{van2016coordinated}
E.~Van~der Pol and F.~A. Oliehoek, ``Coordinated deep reinforcement learners
  for traffic light control,'' in \emph{NIPS'16 Workshop on Learning, Inference
  and Control of Multi-Agent Systems}, Dec. 2016.

\bibitem{kok2006collaborative}
J.~R. Kok and N.~Vlassis, ``Collaborative multiagent reinforcement learning by
  payoff propagation,'' \emph{Journal of Machine Learning Research}, vol.~7,
  pp. 1789--1828, 2006.

\bibitem{bakker2010traffic}
B.~Bakker, S.~Whiteson, L.~Kester, and F.~C. Groen, ``Traffic light control by
  multiagent reinforcement learning systems,'' in \emph{Interactive
  Collaborative Information Systems}.\hskip 1em plus 0.5em minus 0.4em\relax
  Springer, 2010, pp. 475--510.

\bibitem{wiering2000multi}
M.~Wiering, ``Multi-agent reinforcement learning for traffic light control,''
  in \emph{Machine Learning: Proceedings of the Seventeenth International
  Conference (ICML'2000)}, 2000, pp. 1151--1158.

\bibitem{chu2019multi}
T.~Chu, J.~Wang, L.~Codec{\`a}, and Z.~Li, ``Multi-agent deep reinforcement
  learning for large-scale traffic signal control,'' \emph{IEEE Transactions on
  Intelligent Transportation Systems}, vol.~21, pp. 1086--1095, 2019.

\bibitem{presslight19}
H.~Wei, C.~Chen, G.~Zheng, K.~Wu, V.~Gayah, K.~Xu, and Z.~Li, ``Presslight:
  Learning max pressure control to coordinate traffic signals in arterial
  network,'' in \emph{Proceedings of the 25th ACM SIGKDD International
  Conference on Knowledge Discovery \& Data Mining}, 2019, pp. 1290--1298.

\bibitem{varaiya2013max}
P.~Varaiya, ``The max-pressure controller for arbitrary networks of signalized
  intersections,'' in \emph{Advances in Dynamic Network Modeling in Complex
  Transportation Systems}.\hskip 1em plus 0.5em minus 0.4em\relax Springer,
  2013, pp. 27--66.

\bibitem{nishi2018traffic}
T.~Nishi, K.~Otaki, K.~Hayakawa, and T.~Yoshimura, ``Traffic signal control
  based on reinforcement learning with graph convolutional neural nets,'' in
  \emph{2018 21st International Conference on Intelligent Transportation
  Systems (ITSC)}.\hskip 1em plus 0.5em minus 0.4em\relax IEEE, 2018, pp.
  877--883.

\bibitem{colight}
H.~Wei, N.~Xu, H.~Zhang, G.~Zheng, X.~Zang, C.~Chen, W.~Zhang, Y.~Zhu, K.~Xu,
  and Z.~Li, ``Colight: Learning network-level cooperation for traffic signal
  control,'' in \emph{Proceedings of the 28th ACM International Conference on
  Information and Knowledge Management}, 2019, pp. 1913--1922.

\bibitem{bazzi2016distributed}
A.~Bazzi, A.~Zanella, and B.~M. Masini, ``A distributed virtual traffic light
  algorithm exploiting short range {V2V} communications,'' \emph{Ad Hoc
  Networks}, vol.~49, pp. 42--57, 2016.

\bibitem{ferreira2011impact}
M.~Ferreira and P.~M. d'Orey, ``On the impact of virtual traffic lights on
  carbon emissions mitigation,'' \emph{IEEE Transactions on Intelligent
  Transportation Systems}, vol.~13, pp. 284--295, 2011.

\bibitem{gupta2017cooperative}
J.~K. Gupta, M.~Egorov, and M.~Kochenderfer, ``Cooperative multi-agent control
  using deep reinforcement learning,'' in \emph{International Conference on
  Autonomous Agents and Multiagent Systems}.\hskip 1em plus 0.5em minus
  0.4em\relax Springer, 2017, pp. 66--83.

\bibitem{panait2005cooperative}
L.~Panait and S.~Luke, ``Cooperative multi-agent learning: The state of the
  art,'' \emph{Autonomous agents and multi-agent systems}, vol.~11, pp.
  387--434, 2005.

\bibitem{foerster2016learning}
J.~Foerster, I.~A. Assael, N.~de~Freitas, and S.~Whiteson, ``Learning to
  communicate with deep multi-agent reinforcement learning,'' in \emph{Advances
  in Neural Information Processing Systems}, 2016, pp. 2137--2145.

\bibitem{sukhbaatar2016learning}
S.~Sukhbaatar, R.~Fergus \emph{et~al.}, ``Learning multiagent communication
  with backpropagation,'' in \emph{Advances in Neural Information Processing
  Systems}, 2016, pp. 2244--2252.

\bibitem{conti2019soft}
A.~Conti, S.~Mazuelas, S.~Bartoletti, W.~C. Lindsey, and M.~Z. Win, ``Soft
  information for localization-of-things,'' \emph{Proceedings of the IEEE},
  vol. 107, pp. 2240--2264, 2019.

\bibitem{win2018network}
M.~Z. Win, W.~Dai, Y.~Shen, G.~Chrisikos, and H.~V. Poor, ``Network operation
  strategies for efficient localization and navigation,'' \emph{Proceedings of
  the IEEE}, vol. 106, pp. 1224--1254, 2018.

\bibitem{sharma2019decentralized}
P.~Sharma, A.-A. Saucan, D.~J. Bucci, and P.~K. Varshney, ``Decentralized
  gaussian filters for cooperative self-localization and multi-target
  tracking,'' \emph{IEEE Transactions on Signal Processing}, vol.~67, pp.
  5896--5911, 2019.

\bibitem{dunbabin2009experiments}
M.~Dunbabin, P.~Corke, I.~Vasilescu, and D.~Rus, ``Experiments with cooperative
  control of underwater robots,'' \emph{The International Journal of Robotics
  Research}, vol.~28, pp. 815--833, 2009.

\bibitem{ngai2011multiple}
D.~C.~K. Ngai and N.~H.~C. Yung, ``A multiple-goal reinforcement learning
  method for complex vehicle overtaking maneuvers,'' \emph{IEEE Transactions on
  Intelligent Transportation Systems}, vol.~12, pp. 509--522, 2011.

\bibitem{mnih2013playing}
V.~Mnih, K.~Kavukcuoglu, D.~Silver, A.~Graves, I.~Antonoglou, D.~Wierstra, and
  M.~Riedmiller, ``Playing atari with deep reinforcement learning,'' in
  \emph{NIPS Deep Learning Workshop}, 2013.

\bibitem{tampuu2017multiagent}
A.~Tampuu, T.~Matiisen, D.~Kodelja, I.~Kuzovkin, K.~Korjus, J.~Aru, J.~Aru, and
  R.~Vicente, ``Multiagent cooperation and competition with deep reinforcement
  learning,'' \emph{PloS one}, vol.~12, p. e0172395, 2017.

\bibitem{zawadzki2014empirically}
E.~Zawadzki, A.~Lipson, and K.~Leyton-Brown, ``Empirically evaluating
  multiagent learning algorithms,'' \emph{arXiv preprint arXiv:1401.8074},
  2014.

\bibitem{lowe2017multi}
R.~Lowe, Y.~Wu, A.~Tamar, J.~Harb, O.~P. Abbeel, and I.~Mordatch, ``Multi-agent
  actor-critic for mixed cooperative-competitive environments,'' in
  \emph{Advances in Neural Information Processing Systems}, 2017, pp.
  6379--6390.

\bibitem{jiang2018graph}
\BIBentryALTinterwordspacing
J.~Jiang, C.~Dun, T.~Huang, and Z.~Lu, ``Graph convolutional reinforcement
  learning,'' in \emph{8th International Conference on Learning
  Representations, {ICLR} 2020, Addis Ababa, Ethiopia, April 26-30,
  2020}.\hskip 1em plus 0.5em minus 0.4em\relax OpenReview.net, 2020. [Online].
  Available: \url{https://openreview.net/forum?id=HkxdQkSYDB}
\BIBentrySTDinterwordspacing

\bibitem{scarselli2009graph}
F.~Scarselli, M.~Gori, A.~C. Tsoi, M.~Hagenbuchner, and G.~Monfardini, ``The
  graph neural network model,'' \emph{IEEE Transactions on Neural Networks},
  vol.~20, pp. 61--80, 2009.

\bibitem{battaglia2018relational}
P.~W. Battaglia, J.~B. Hamrick, V.~Bapst, A.~Sanchez-Gonzalez, V.~Zambaldi,
  M.~Malinowski, A.~Tacchetti, D.~Raposo, A.~Santoro, R.~Faulkner
  \emph{et~al.}, ``Relational inductive biases, deep learning, and graph
  networks,'' \emph{arXiv preprint arXiv:1806.01261}, 2018.

\bibitem{battaglia2016interaction}
P.~Battaglia, R.~Pascanu, M.~Lai, D.~J. Rezende \emph{et~al.}, ``Interaction
  networks for learning about objects, relations and physics,'' in
  \emph{Advances in neural information processing systems}, 2016, pp.
  4502--4510.

\bibitem{li2015gated}
\BIBentryALTinterwordspacing
Y.~Li, D.~Tarlow, M.~Brockschmidt, and R.~S. Zemel, ``Gated graph sequence
  neural networks,'' in \emph{4th International Conference on Learning
  Representations, {ICLR} 2016, San Juan, Puerto Rico, May 2-4, 2016,
  Conference Track Proceedings}, 2016. [Online]. Available:
  \url{http://arxiv.org/abs/1511.05493}
\BIBentrySTDinterwordspacing

\bibitem{zhi2019abductive}
Z.~Zhi-Hua, ``Abductive learning: towards bridging machine learning and logical
  reasoning,'' \emph{Science China (Information Sciences)}, no.~7, p.~21, 2019.

\bibitem{gilmer2017neural}
J.~Gilmer, S.~S. Schoenholz, P.~F. Riley, O.~Vinyals, and G.~E. Dahl, ``Neural
  message passing for quantum chemistry,'' in \emph{Proceedings of the 34th
  International Conference on Machine Learning, {ICML}}, 2017, pp. 1263--1272.

\bibitem{wang2018nervenet}
\BIBentryALTinterwordspacing
T.~Wang, R.~Liao, J.~Ba, and S.~Fidler, ``Nervenet: Learning structured policy
  with graph neural networks,'' in \emph{6th International Conference on
  Learning Representations, {ICLR} 2018, Vancouver, BC, Canada, April 30 - May
  3, 2018, Conference Track Proceedings}.\hskip 1em plus 0.5em minus
  0.4em\relax OpenReview.net, 2018. [Online]. Available:
  \url{https://openreview.net/forum?id=S1sqHMZCb}
\BIBentrySTDinterwordspacing

\bibitem{zambaldi2018relational}
\BIBentryALTinterwordspacing
V.~F. Zambaldi, D.~Raposo, A.~Santoro, V.~Bapst, Y.~Li, I.~Babuschkin,
  K.~Tuyls, D.~P. Reichert, T.~P. Lillicrap, E.~Lockhart, M.~Shanahan,
  V.~Langston, R.~Pascanu, M.~Botvinick, O.~Vinyals, and P.~W. Battaglia,
  ``Deep reinforcement learning with relational inductive biases,'' in
  \emph{7th International Conference on Learning Representations, {ICLR} 2019,
  New Orleans, LA, USA, May 6-9, 2019}.\hskip 1em plus 0.5em minus 0.4em\relax
  OpenReview.net, 2019. [Online]. Available:
  \url{https://openreview.net/forum?id=HkxaFoC9KQ}
\BIBentrySTDinterwordspacing

\bibitem{wiering2004simulation}
M.~Wiering, J.~Vreeken, J.~Van~Veenen, and A.~Koopman, ``Simulation and
  optimization of traffic in a city,'' in \emph{IEEE Intelligent Vehicles
  Symposium, 2004}.\hskip 1em plus 0.5em minus 0.4em\relax IEEE, 2004, pp.
  453--458.

\bibitem{hochreiter1997long}
S.~Hochreiter and J.~Schmidhuber, ``Long short-term memory,'' \emph{Neural
  computation}, vol.~9, pp. 1735--1780, 1997.

\bibitem{tan1993multi}
M.~Tan, ``Multi-agent reinforcement learning: Independent vs. cooperative
  agents,'' in \emph{Proceedings of the tenth international conference on
  machine learning}, 1993, pp. 330--337.

\bibitem{xu2015empirical}
B.~Xu, N.~Wang, T.~Chen, and M.~Li, ``Empirical evaluation of rectified
  activations in convolutional network,'' \emph{arXiv preprint
  arXiv:1505.00853}, 2015.

\bibitem{vaswani2017attention}
A.~Vaswani, N.~Shazeer, N.~Parmar, J.~Uszkoreit, L.~Jones, A.~N. Gomez,
  {\L}.~Kaiser, and I.~Polosukhin, ``Attention is all you need,'' in
  \emph{Advances in Neural Information Processing Systems}, 2017, pp.
  5998--6008.

\bibitem{clevert2015fast}
\BIBentryALTinterwordspacing
D.~Clevert, T.~Unterthiner, and S.~Hochreiter, ``Fast and accurate deep network
  learning by exponential linear units (elus),'' in \emph{4th International
  Conference on Learning Representations, {ICLR} 2016, San Juan, Puerto Rico,
  May 2-4, 2016, Conference Track Proceedings}, Y.~Bengio and Y.~LeCun, Eds.,
  2016. [Online]. Available: \url{http://arxiv.org/abs/1511.07289}
\BIBentrySTDinterwordspacing

\bibitem{hausknecht2015deep}
M.~J. Hausknecht and P.~Stone, ``Deep recurrent q-learning for partially
  observable mdps,'' in \emph{2015 {AAAI} Fall Symposia, Arlington, Virginia,
  USA, November 12-14, 2015}.\hskip 1em plus 0.5em minus 0.4em\relax {AAAI}
  Press, 2015, pp. 29--37.

\bibitem{heess2015memory}
N.~Heess, J.~J. Hunt, T.~P. Lillicrap, and D.~Silver, ``Memory-based control
  with recurrent neural networks,'' in \emph{NIPS Deep Reinforcement Learning
  Workshop}, 2015.

\bibitem{mnih2015human}
V.~Mnih, K.~Kavukcuoglu, D.~Silver, A.~A. Rusu, J.~Veness, M.~G. Bellemare,
  A.~Graves, M.~Riedmiller, A.~K. Fidjeland, G.~Ostrovski \emph{et~al.},
  ``Human-level control through deep reinforcement learning,'' \emph{Nature},
  vol. 518, p. 529, 2015.

\bibitem{zhang2019cityflow}
H.~Zhang, S.~Feng, C.~Liu, Y.~Ding, Y.~Zhu, Z.~Zhou, W.~Zhang, Y.~Yu, H.~Jin,
  and Z.~Li, ``Cityflow: A multi-agent reinforcement learning environment for
  large scale city traffic scenario,'' in \emph{The World Wide Web
  Conference}.\hskip 1em plus 0.5em minus 0.4em\relax ACM, 2019, pp.
  3620--3624.

\bibitem{he2015delving}
K.~He, X.~Zhang, S.~Ren, and J.~Sun, ``Delving deep into rectifiers: Surpassing
  human-level performance on imagenet classification,'' in \emph{Proceedings of
  the IEEE international conference on computer vision}, 2015, pp. 1026--1034.

\bibitem{kingma2014adam}
\BIBentryALTinterwordspacing
D.~P. Kingma and J.~Ba, ``Adam: {A} method for stochastic optimization,'' in
  \emph{3rd International Conference on Learning Representations, {ICLR} 2015,
  San Diego, CA, USA, May 7-9, 2015, Conference Track Proceedings}, 2015.
  [Online]. Available: \url{http://arxiv.org/abs/1412.6980}
\BIBentrySTDinterwordspacing

\end{thebibliography}


\end{document}